\documentclass[12pt]{article}
\usepackage{amsmath,amssymb,bm,epsfig,color,graphicx}
\renewcommand{\thefootnote}{\fnsymbol{footnote}}

\begin{document}

\title{
\begin{flushright}
\begin{minipage}{0.2\linewidth}
\normalsize
WU-HEP-16-15 \\*[50pt]
\end{minipage}
\end{flushright}
{\Large \bf 
Axion decay constants at special points\\ in type II string theory
\\*[20pt]}}

\author{Masaki~Honda$^{1,}$\footnote{
E-mail address: yakkuru$\_$111@ruri.waseda.jp},\ \ 
Akane~Oikawa$^{1,}$\footnote{
E-mail address: a.oikawa@aoni.waseda.jp}, \ and \
Hajime~Otsuka$^{1,}$\footnote{
E-mail address: h.otsuka@aoni.waseda.jp}
\\*[20pt]
$^1${\it \normalsize 
Department of Physics, Waseda University, 
Tokyo 169-8555, Japan}  \\*[50pt]}

\date{
\centerline{\small \bf Abstract}
\begin{minipage}{0.9\linewidth}
\medskip 
\medskip 
\small
We propose the mechanism to disentangle the decay constant of closed string axion from the string scale 
in the framework of type II string theory on Calabi-Yau manifold. 
We find that the quantum and geometrical corrections in the prepotential that arise at some special points in the moduli space widen the window of axion decay constant. 
In particular, around the small complex structure points, 
the axion decay constant becomes significantly lower than the string scale. 
We also discuss the moduli stabilization leading to the phenomenologically attractive low-scale axion decay constant. 
\end{minipage}
}

\begin{titlepage}
\maketitle
\thispagestyle{empty}
\clearpage
\end{titlepage}

\renewcommand{\thefootnote}{\arabic{footnote}}
\setcounter{footnote}{0}

\section{Introduction}
An axion, or axion-like particle, is phenomenologically 
and cosmologically attractive particle 
to explain the origin of tiny strong CP phase in the standard model~\cite{Peccei:1977hh}, 
the current dark matter abundance~\cite{Preskill:1982cy,Abbott:1982af,Dine:1982ah} and the origin of 
cosmological microwave background through the inflation mechanism. 
The consistent theory of quantum gravity such as the string theory 
also predicts the existence of axion particles through the dimensional 
reduction of the higher-dimensional vector and tensor fields associated with 
the internal cycles of extra-dimensional space.

In particular, the QCD axion decay constant should be within the range, $10^{9-12}$ GeV 
by the observation of supernova (SN) 1987A~\cite{Hirata:1987hu,Bionta:1987qt} and dark matter 
abundance observed by Planck~\cite{Ade:2015lrj} with ${\cal O}(1)$ initial misalignment angle.(For a 
review, see, e.g., Ref.~\cite{Kim:1986ax}.) 
From the inflationary point of view, the decay constant of axion inflaton 
is severely constrained by the Planck data~\cite{Ade:2015lrj}, e.g., $10^{18-19}$ GeV for 
the natural inflation~\cite{Freese:1990rb}. 
It is difficult to extract the above constrained axion decay constant from the 
string theory. 
The authors of Refs.~\cite{Choi:1985je,Banks:2003sx,Svrcek:2006yi} showed that 
decay constant of closed string axion in the string theory 
is typically around $10^{16-17}$ GeV, since both the decay constant and the gauge coupling of 
visible sector are closely related through the volume of extra-dimensional space. 
When the visible sector lives on the localized cycle of extra-dimensional space, 
the decay constant of axion associated with the large volume cycle can be taken 
much smaller than Planck scale~\cite{Conlon:2006tq}, which is achieved in so-called LARGE Volume Scenario (LVS) in 
type IIB string theory~\cite{Balasubramanian:2005zx}.\footnote{The warped string compactification is also discussed 
in Ref.~\cite{Dasgupta:2008hb}. See for other axions irrelevant with the internal cycle of extra-dimensional 
space in heterotic string~\cite{Buchbinder:2014qca} and type IIA string~\cite{Honecker:2013mya}.} 
On the other hand, the larger axion decay constant is obtained by the inclusion of 
the quantum corrections to the volume of extra-dimensional space~\cite{Conlon:2016aea}, 
one-loop corrections for the gauge couplings~\cite{Abe:2014pwa,Abe:2014xja}, and 
the alignment mechanism in terms of multiple axions~\cite{Kim:2004rp}.

In this paper, we focus on K\"ahler (complex structure) moduli fields 
in type IIA (IIB) string theory on Calabi-Yau (CY) manifold. 
These moduli potentials receive the quantum (geometrical) corrections which are 
exactly calculated in the topological string theory. 
Recently, the authors of Ref.~\cite{Conlon:2016aea} have showed that the axion decay constant including 
the instanton corrections has the maximum value in type IIA string on CY manifold with a few moduli fields 
around the large volume limit, which corresponds to the large complex structure (LCS) limit in type IIB string on 
mirror CY manifold. 
In this case, even if the volume of internal cycle is of ${\cal O}(1)$ in string units, 
such instanton corrections give the sizable effect. 
In this way, it motivates us to proceed to study the detail of the quantum and geometrical corrections 
for a decay constant of closed string axion around the several points of moduli spaces. 
In type IIB string theory on CY manifold, these geometrical and non-perturbative corrections are exactly obtained by 
solving the corresponding Picard-Fuchs differential equation for the period vector of CY manifold, 
which corresponds to the instanton corrections in type IIA side. 

We in particular focus on the regular singular points involved in the Picard-Fuchs equation which are called as 
the special points of moduli space, such as the LCS point, conifold point, 
and small complex structure (SCS) point involving the Gepner point~\cite{Gepner:1987vz,Gepner:1987qi}. 
These closed string axions then naturally appear around the special points 
in the low-energy effective theory, since the 
monodromy symmetries around special points allow the existence of axions in the moduli K\"ahler potential. 
Around these special points, we proceed to study the detail of the quantum and 
geometrical corrections for a decay constant of closed string axion.
It is remarkable that in type IIB string theory, the decay constants of axions associated with the complex structure moduli are irrelevant to the volume of mirror CY manifold, i.e., the string scale, in comparison 
with those of K\"ahler moduli. 
As pointed out in Refs.~\cite{Garcia-Etxebarria:2014wla,Kobayashi:2015aaa,Akita:2016usy}, it is interesting to discuss the phenomenology and cosmology of 
complex structure moduli.

The remaining of this paper is organized as follows. 
After briefly reviewing the K\"ahler potential on the basis of ${\cal N}=2$ special geometry in 
Sec.~\ref{sec:2}, we first show how to define the axion particles around the special points of complex structure 
moduli space and corresponding decay constant. 
The axionic shift symmetries are then captured by the invariance of K\"ahler 
potential under the monodromy transformation at the special points. 
In Sec.~\ref{subsec:scs}, we formulate the geometrical corrections for the period vector of CY manifold 
with an emphasis on the SCS point. 
It is then found that, in contrast to the previous studies, the decay 
constant of closed string axion associated with the complex structure modulus 
is taken much smaller than the string scale, and 
such a property is a common phenomena in typical one-parameter CY manifolds. 
Next, we proceed to estimate the axion decay constant around 
the conifold point in Sec.~\ref{subsec:coni} and LCS point in Sec.~\ref{subsec:LCS}. 
Finally, we comment on the moduli stabilization to generate 
the low-scale axion decay constants in Sec.~\ref{sec:moduli}. 
Sec.~\ref{sec:con} is devoted to the conclusion.

\section{Decay constant of closed string axion}
\label{sec:2}
Throughout this paper, we consider the low-energy effective theory of 
type II string theory on CY orientifold in which the closed string moduli 
are categorized into the axion-dilaton $\tau$, K\"ahler moduli $T$, and 
complex structure moduli $z$.(See for the construction of 
four-dimensional ${\cal N}=1$ effective theory on CY orientifold, e.g., Ref.~\cite{Grimm:2004uq}.)

First of all, we show the decay constant of axion associated with 
K\"ahler moduli in type IIB string theory on CY orientifold.\footnote{It is straightforward to extend 
the following discussion to type IIA string theory or 
heterotic string with standard embedding.} 
In the framework of four-dimensional ${\cal N}=1$ supergravity action, 
the relevant Lagrangian density of closed string axion associated with the 
Kalb-Ramond field or Ramond-Ramond field is captured by
\begin{align}
{\cal L}=-K_{\rho \bar{\rho}} \partial_\mu \theta\partial^\mu \theta 
-\frac{1}{4g^2}F^{\mu\nu}F_{\mu\nu}-\frac{\theta}{4}F^{\mu\nu}\tilde{F}_{\mu\nu},
\end{align}
where $\theta(x)$ is the axion accompanied by the certain modulus field $\rho$, 
$K_{\rho \bar{\rho}}=\partial^2 K/\partial \rho\partial \bar{\rho}$ is the second derivative of the 
K\"ahler potential $K$ with respect to $\rho$ and $\bar{\rho}$, 
$\mu, \nu$ denote the indices of 
Minkowski spacetime, and $\tilde{F}_{\mu\nu}$ is the dual gauge field strength of $F_{\mu\nu}$ 
in the visible or hidden sector with the gauge coupling $g$. 
Since the axion potential is non-perturbatively generated by breaking the higher-dimensional 
gauge symmetries to discrete one, the axion decay constant $f$ is extracted by canonically 
normalizing the axion as $\tilde{\theta}=\sqrt{2K_{\rho\bar{\rho}}}\theta$,
\begin{align}
{\cal L}=-\frac{1}{2}\partial_\mu \tilde{\theta}\partial^\mu \tilde{\theta} 
-\frac{\tilde{\theta}}{32\pi^2f}F^{\mu\nu}\tilde{F}_{\mu\nu},
\end{align}
with 
\begin{align}
f= \frac{\sqrt{2K_{\rho \bar{\rho}}}}{8\pi^2}.
\end{align}
When $\rho$ corresponds the K\"ahler modulus $T$ of CY manifold ${\cal M}$ in the type IIB string theory, 
the classical modulus K\"ahler potential is characterized 
by the volume of CY manifold ${\cal V}$ in string units, i.e., 
$K=-2M_{\rm Pl}^2\ln {\cal V}$ with $M_{\rm Pl}$ being the reduced Planck mass\footnote{We use the reduced Planck unit 
$M_{\rm Pl}=2.4\times 10^{18}{\rm GeV}=1$ unless otherwise specified.}. 
This K\"ahler potential is valid in the large volume limit of CY manifold 
${\cal V}=(T+\bar{T})^{3/2}> 1$, and then the decay constant 
of axion ${\rm Im}\,T$ reads as
\begin{align}
f\simeq \frac{\sqrt{6}}{{\cal V}^{2/3}}\frac{M_{\rm Pl}}{8\pi^2}.
\end{align}
For the closed string axion in the string theory, 
the (internal) volume of CY manifold is related to the value of gauge coupling. 
Thus, the observed value of gauge coupling in the visible sector leads to the constrained 
axion decay constant around $10^{16-17}$ GeV~\cite{Svrcek:2006yi}, 
which is the same order of the string scale 
\begin{align}
M_{s}\simeq \frac{g_s}{\sqrt{4\pi {\cal V}}}M_{\rm Pl},
\label{eq:stscale}
\end{align}
with $g_s$ being the string coupling. 
The similar discussion is applied to the case of axion-dilaton $\tau$. 
By contrast, for the internal cycle irrelevant to the visible sector, 
tiny axion decay constant is achieved in the large volume limit, in particular, 
phenomenologically favorable axion window, $10^9\leq f\leq 10^{12}$ GeV corresponds to the range, 
$2\times 10^{7}\leq {\cal V} \leq 6.4\times 10^{11}$. Furthermore, when we consider the hidden D$7$-brane 
wrapping the small cycle in ``Swiss-Cheese'' CY manifold, the corresponding axion decay constant 
can be taken much smaller than the string scale independently of the overall volume of CY manifold. 
In both cases, the string scale is simultaneously small compared with the Planck scale. 
On the other hand, the trans-Planckian axion decay constant requires the small volume of CY manifold. 
When the CY volume is of order unity in string units, the quantum corrections will give an important role of 
determining the axion decay constant as suggested in Ref.~\cite{Conlon:2016aea}. 

In the following, let us take a closer look at the complex structure moduli in type IIB string theory. 
In contrast to the K\"ahler moduli in type IIB string, 
the vacuum expectation values of complex structure moduli are irrelevant to the value of string scale 
as can be seen in Eq.~(\ref{eq:stscale}). 
It is thus expected that the decay constant of axion associated with complex structure moduli 
is decoupled from the string scale. 
Furthermore, 
since the geometrical and non-perturbative corrections for the complex structure 
moduli are exactly calculated in the topological string theory, we can estimate such corrections 
to the axion decay constant around several points of moduli space in a systematic way. 
Note that, in type IIB (IIA) string theory on toroidal background, 
axions associated with complex structure (K\"ahler) moduli couple to the gauge bosons 
at the one-loop level through the gauge threshold corrections~\cite{Lust:2003ky,Blumenhagen:2006ci}, 
although such gauge threshold corrections are unrevealed for the CY background. 
These one-loop corrections are induced by integrating out the massive open strings 
between two parallel D-branes. 
Thus, in a way similar to the toroidal background~\cite{Lust:2003ky,Blumenhagen:2006ci}, 
we assume that the one-loop gauge kinetic function involves the axion associated 
with complex structure moduli so that the gauge couplings do not diverge 
around the special points. 

Before going to the detail of such corrections in the moduli K\"ahler potential, 
we briefly review the K\"ahler potential written in the language of ${\cal N}=2$ special geometry. 
In type II string theory on CY manifold, the K\"ahler potential originating from ${\cal N}=2$ vector 
multiplets is provided by the prepotential, 
which receives the quantum and geometric corrections, calculated by the technique of mirror symmetry 
between A- and B-models~\cite{Witten:1988xj,Witten:1991zz}. 
For sake of simplicity, we consider the type IIB string theory on CY orientifold. 
On the integral symplectic cohomology basis for $H^{3}({\cal M}, \mathbb{Z})$ of CY manifold, ($\alpha_a, \beta^a$) with 
$a=1,2,\cdots, h^{2,1}+1$, where $h^{2,1}$ denote the hodge number surviving under the orientifold 
involution, 
the holomorphic three-form is represented by
\begin{align}
\Omega =z^a \alpha_a -{\cal F}_a \beta^a,
\end{align}
where $z^a$ and ${\cal F}_a=\partial {\cal F}/\partial z^a$ are defined 
by the integrals of $\Omega$ over three-cycles $A_a$ and $B^a$ on the integral homology basis 
in $H_3({\cal M}, \mathbb{Z})$, 
\begin{align}
z^a =\int_{A_a} \Omega,\nonumber\\
{\cal F}_a =\int_{B^a} \Omega, 
\end{align}
respectively. 
Then, ${\cal N}=2$ special geometry allows us to write down the 
K\"ahler potential of complex structure moduli,
\begin{align}
K=-\ln \bigl[ i\int_{\cal M} \Omega \wedge \bar{\Omega}\bigl]
=-\ln \bigl[ i\Pi^\dagger \cdot \Sigma \cdot \Pi\bigl],
\label{eq:Kahler}
\end{align}
where
\begin{align}
\Pi^t =\left(\int_{A_a} \Omega, \int_{B^a} \Omega \right),
\end{align}
is the period vector and $\Sigma$ is the symplectic matrix:
\begin{align}
\Sigma=
\begin{pmatrix}
0 & {\bm 1}\\
-{\bm 1} & 0\\
\end{pmatrix}
,
\end{align}
with ${\bm 1}$ being the $(h^{2,1}+1)\times (h^{2,1}+1)$ identity matrix. 
When the three-form fluxes are inserted in these three-cycles, the flux-induced 
superpotential is generated as~\cite{Gukov:1999ya}
\begin{align}
W=\int_{\cal M} G_3 \wedge \Omega,
\end{align}
where $G_3=F_3-\tau H_3$ denote the linear combination of Ramond-Ramond $F_3$ 
and Neveu-Schwarz three-form fluxes $H_3$. 
These three-form fluxes enable us to stabilize the complex structure moduli and axion-dilaton~\cite{Giddings:2001yu}. 
It is remarkable that the K\"ahler potential is invariant under the symplectic transformation:
\begin{align}
\Pi^i(z) \rightarrow \sum_{j}P^i_j \Pi^j(z),
\end{align}
where the integral symplectic transformation matrix $P^i_j$, ($i,j=1,2,\cdots, 2(h^{2,1}+1)$), is the 
matrix representation of symplectic group $Sp(2(h^{2,1}+1))$. 

In the complex structure moduli space, there are several special points 
called as SCS point, conifold point, and LCS point, 
around which the period vector transforms under the discrete subgroup (monodromy group) of 
symplectic group, 
\begin{align}
\sum_{j}P^i_j[z_{\rm sp}] \Pi^j(z)=\Pi^i(ze^{2\pi i}),
\end{align}
with $P^i_j[z_{\rm sp}]$ being the monodromy matrix around the 
special point $z_{\rm sp}$. 
Such symmetries of the K\"ahler potential around the special points 
give rise to the flat direction in moduli space. 
Thus, an axion or axion-like particle naturally appears in the low-energy effective theory. 

To estimate the axion decay constant around the special points, 
it is required to derive an exact form of period vector and its monodromy transformation. 
In the remaining of this paper, we show the systematic approach to find the general expression of period vector 
and its monodromy transformation for one-parameter CY 
manifold within the framework of topological string theory, in which these period vectors are governed by 
the Picard-Fuchs differential equations. 
It thus allows us to study the axion decay constant  around the special points of moduli space.

\section{Axion decay constant around the special point}
\label{sec:3}
We discuss the decay constant of axion around the 
special points of the complex structure moduli space step by step. 
First of all, we take into account the SCS points for several CY manifolds in Sec.~\ref{subsec:scs}.\footnote{In what 
follows, the mirror CY manifold is denoted by the CY manifold for simplicity, unless otherwise specified.}
In Sec.~\ref{subsec:coni}, we discuss the conifold point of CY manifold, in particular, mirror quintic 
CY threefold. Finally, the LCS (or large volume) point is discussed in Sec.~\ref{subsec:LCS}.

\subsection{SCS point}
\label{subsec:scs}
To make the analysis concrete, we, in this section, 
restrict ourselves to the type IIB string theory on mirror CY orientifold 
with one complex structure modulus. 
Furthermore, we concentrate on the complex structure modulus 
compatible with the orientifold involution.(See for the construction of 
four-dimensional ${\cal N}=1$ effective theory on CY orientifold, e.g., Ref.~\cite{Grimm:2004uq}.) 
In particular, when a CY threefold is defined as a hypersurface in 
a weighted projective space $\mathbb{P}^n_{\nu_1,\nu_2,\cdots, \nu_{n+1}}[d_1,d_2,\cdots, d_r]$, 
where the degree of homogeneous polynomials $d_1,d_2,\cdots,d_r$ 
satisfies the so-called CY condition: $\sum_{p=1}^r d_p=\sum_{q=1}^{n+1}\nu_q$ with $n=r+3$, 
the period vector obeys the Picard-Fuchs equation~\cite{Hosono:1993qy,Hosono:1994ax}
\begin{align}
\left\{\delta^4
-h z(\delta +\alpha_1)(\delta +\alpha_2)(\delta +\alpha_3)(\delta +\alpha_4)\right\}
\pi (z)=0,
\label{eq:PF}
\end{align}
where $\delta=zd/dz$ with $z$ being the complex structure modulus, 
$h=\Pi_p d_p^{d_p}/\Pi_q \nu_q^{\nu_q}$ and 
$\cup_{i=1}^4\{\alpha_i\}=\{ \cup_{p=1}^r\{\frac{l}{d_p}\}\backslash \cup_{q=1}^{n+1}\{\frac{m}{\nu_q}\} |
0\leq l\leq d_p-1, 0\leq m\leq \nu_q-1\}$ for $l,m\in \mathbb{Z}$. The rational numbers $\alpha_i$ characterize the CY manifold. 
Throughout this section, we study the CY manifolds with one complex structure modulus 
defined in complete intersections in weighted projective spaces and toric varieties. 
Such CY manifolds are summarized in Table~\ref{tab1} 
in which those are classified by the degeneracies of $\alpha_i$. 
In Sec.~\ref{subsubsec:orb}, we discuss the case where all $\alpha_i$ are distinct, i.e., 
$\alpha_1 \neq \alpha_2 \neq \alpha_3 \neq \alpha_4$, while the other degenerated cases 
are shown in Secs.~\ref{subsubsec:hyb1},~\ref{subsubsec:hyb2} and~\ref{subsubsec:hyb3}.\footnote{As 
pointed out in Ref.~\cite{Greene:2000ci}, the case with $\alpha_1 = \alpha_2 = \alpha_3 \neq \alpha_4$ will not 
be realized in toric varieties.}
\begin{table}[htb]
\begin{center}
  \begin{tabular}{|c|c|c|c|c|c|} \hline
    CY hypersurfaces & $(\alpha_1, \alpha_2, \alpha_3, \alpha_4)$ & $h$ & $\kappa$ & $\int_{\cal M} c_2\wedge D$ & Degeneracies of $\alpha_i$ 
    \\ \hline \hline
     $\mathbb{P}^4_{1,1,1,1,1}[5]$ 
     & $(1/5,2/5,3/5,4/5)$ & $3125$ & $5$ & $50$ 
     & $0$ \\
     $\mathbb{P}^4_{2,1,1,1,1}[6]$ 
      & $(1/6,2/6,4/6,5/6)$ & $11664$ & $3$ & $42$
     & $0$\\
    $\mathbb{P}^4_{4,1,1,1,1}[8]$ 
     & $(1/8,3/8,5/8,7/8)$ & $65536$ & $2$ & $44$ 
    & $0$\\
    $\mathbb{P}^4_{5,2,1,1,1}[10]$ 
     & $(1/10,3/10,7/10,9/10)$ & $8\cdot 10^5$ & $1$ & $34$
    & $0$\\ 
     $\mathbb{P}^5_{6,4,1,1,1,1}[2,12]$ 
     & $(1/12,5/12,7/12,11/12)$ & $12^6$ & $1$ & $46$
    & $0$\\ 
     $\mathbb{P}^5_{3,2,2,1,1,1}[4,6]$ 
     & $(1/4,1/3,2/3,3/4)$ & $27648$  & $2$ & $32$
    & $0$\\     
     $\mathbb{P}^5_{2,1,1,1,1,1}[3,4]$ 
     & $(1/6,1/4,3/4,5/6)$ & $1728$ & $6$ & $48$
    & $0$\\ 
     \hline
    $\mathbb{P}^5_{1,1,1,1,1,1}[2,4]$ & $(1/4,3/4,1/2,1/2)$ 
    & $1024$ & $8$ & $56$ & 1\\ 
     $\mathbb{P}^6_{1,1,1,1,1,1,1}[2,2,3]$ & $(1/3,2/3,1/2,1/2)$ 
     & $432$ & $12$ & $60$ & $1$\\ 
   $\mathbb{P}^5_{3,1,1,1,1,1}[2,6]$ & $(1/6,5/6,1/2,1/2)$ 
   & $6912$ & $4$ & $52$ & $1$\\ 
    \hline
    $\mathbb{P}^5_{1,1,1,1,1,1}[3,3]$ & $(1/3,1/3,2/3,2/3)$ 
    & $729$ & $9$ & $54$ & $2$\\ 
     $\mathbb{P}^5_{2,2,1,1,1,1}[4,4]$ & $(1/4,1/4,3/4,3/4)$ 
     & $4096$ & $4$ & $40$ & $2$\\ 
   $\mathbb{P}^5_{3,3,2,2,1,1}[6,6]$ & $(1/6,1/6,5/6,5/6)$ 
   & $2^8\cdot 3^6$ & $1$ & $22$ & $2$\\ 
       \hline
    $\mathbb{P}^7_{1,1,1,1,1,1,1,1}[2,2,2,2]$ & $(1/2,1/2,1/2,1/2)$
    & $256$ & $16$ & $64$ & $4$ 
    \\\hline 
    \end{tabular}
    \end{center}
    \caption{The list of the CY threefolds defined in the ambient spaces~\cite{Chen}$^6$, 
    which is classified by the degeneracies of $\alpha_i$ with $i=1,2,3,4$. 
    $h$, $\kappa$, $c_2$ and $D$ denote the constant, triple intersection number, the second Chern 
    class of the tangent bundle, and a divisor on the integral basis of CY manifold, respectively. 
    The derivation of $\alpha_k$ is shown in Ref.~\cite{Hosono:1994ax} for the case of weighted complete intersection.}
    \label{tab1}
\end{table}
\addtocounter{footnote}{1}
\footnotetext{In Ref.~\cite{Chen}, there is a typo in the CY data. Specifically, the CY data of $\mathbb{P}^5_{3,2,2,1,1,1}[4,6]$ needs to be replaced by 
that of $\mathbb{P}^5_{2,1,1,1,1,1}[3,4]$.}

The Picard-Fuchs equation involves the regular singular points, i.e., special points, 
such as the small complex structure point $z= \infty$, conifold 
point $z= 1/h$ and large complex structure point $z= 0$. 
By solving the above Picard-Fuchs equation around the small complex 
structure point of CY manifold $z\rightarrow \infty$, 
the monodromy group is mainly categorized into two classes. 
Firstly, the monodromy matrix is 
of finite order, i.e., $P[z_{\rm sp}]^n=P[z_{\rm sp}]$, where $n$ 
is the integer depending on the structure of CY manifold. 
In such a case, all $\alpha_k$ are distinct and we will call such a special point 
as orbifold point as discussed in detail in Sec.~\ref{subsubsec:orb}. 
Secondly, the monodromy matrix does 
not obey $P[z_{\rm sp}]^n= P[z_{\rm sp}]$ ($n\in \mathbb{Z}$), 
while it satisfies $(P[z_{\rm sp}]^n-1)^m= 0$ for certain integers $n$ and $m$. 
In such a case, some (all) of $\alpha_k$ are degenerate and the period vector has a 
logarithmic behavior as displayed in Secs.~\ref{subsubsec:hyb1},~\ref{subsubsec:hyb2} and \ref{subsubsec:hyb3}.

First of all, we show the general expression of period vector for the CY manifold with 
degeneracies of $\alpha_i$ less than $2$. 
The solution of Picard-Fuchs equation can be written 
in terms of the integral representation of Meijer $G$-function~\cite{Greene:2000ci},(For details about 
the Meijer $G$-function, see, e.g., Refs.~\cite{Greene:2000ci,Gfunction}.) 
\begin{align}
\pi(z)=
\begin{pmatrix}
U_0(z)\\
U_1(z)\\
U_2(z)\\
U_3(z)\\
\end{pmatrix}
,
\label{eq:piJ}
\end{align}
where 
\begin{align}
U_j(z)=\frac{1}{(2\pi i)^{j+1}} \int_C ds \frac{\Gamma (-s)^{j+1}\Pi_{i=1}^4 \Gamma ( s+\alpha_i)}{\Gamma (s+1)^{3-j}\Pi_{i=1}^4 \Gamma (\alpha_i)}
\left( e^{\pi i(j+1)} hz\right)^s,
\label{eq:Uj}
\end{align}
with $j=0,1,2,3$. 
In the SCS point, 
the contour $C$ is taken to extend from $-i\infty$ to $i\infty$ so as to
enclose the poles $s=-\alpha_i-n$ with $n$ being non-negative integer. 

Let us calculate the period integral involving the $n_s$ single poles and $n_d$ double poles. 
Although we do not consider the case of triple poles throughout this paper, quadruple pole is discussed 
in Sec.~\ref{subsubsec:hyb3}. 
The general expression consists of two parts:
\begin{align}
U_j(z) =U_j(z)^{(s)} +U_j(z)^{(d)},
\end{align}
where $U_j(z)^{(s)}$ is the contribution from the single poles at $s=-\alpha_k-n$
\begin{align}
U_j(z)^{(s)}=&\frac{1}{(2\pi i)^{j}} 
\sum_{k}^{n_s} \left(\frac{{\rm sin}(\pi \alpha_k)}{\pi}\right)^{3-j} (e^{\pi i(j+1)}h z)^{-\alpha_k}
\frac{\Gamma(\alpha_k)^4 \Pi_{i=1,i\neq k}^4\Gamma (\alpha_i-\alpha_k)}{\Gamma (\alpha_1)\Gamma (\alpha_2)\Gamma (\alpha_3)\Gamma (\alpha_4)} 
\nonumber\\
&\cdot\sum_{n=0}^\infty \frac{[ (\alpha_k)_n]^4}{n! \Pi_{l=1,\cdots,4,l\neq k} (1+\alpha_k-\alpha_l)_n} (hz)^{-n},
\label{eq:Ujs}
\end{align}
with $(\alpha_k)_n=\frac{\Gamma(\alpha_k +n)}{\Gamma(\alpha_k)}$ being the Pochhammer symbol 
and the second part $U_j(z)^{(d)}$ is originating from the double poles at $s=-\alpha_k-n$ depending on 
the logarithmic term
\begin{align}
U_j(z)^{(d)}=&\frac{1}{(2\pi i)^{j}} 
\sum_{k}^{n_d}
\left(\frac{{\rm sin}(\pi \alpha_k)}{\pi}\right)^{3-j} (e^{\pi i(j+1)} hz)^{-\alpha_k}
\sum_{n=0}^\infty \frac{\Gamma(\alpha_k +n)^4 \Pi_{i=1,i\neq k, \alpha_i\neq\alpha_k}^4
\Gamma (\alpha_i-\alpha_k-n)}{n!^2\Gamma (\alpha_1)\Gamma (\alpha_2)\Gamma (\alpha_3)\Gamma (\alpha_4)} 
\nonumber\\
&\cdot\biggl[ B_k^n-(3-j)\pi{\rm cot}(\pi (\alpha_k+n))+\ln \left(e^{\pi i(j+1)} hz\right)\biggl](hz)^{-n},
\label{eq:Ujd}
\end{align}
with $B_k^n=\sum_{i=1,i\neq k, \alpha_i\neq \alpha_k}^4
\Psi (\alpha_i-\alpha_k-n)-4\Psi (\alpha_k +n) -2\gamma +2\sum_{l=1}^{n} \frac{1}{l}$. 
Here, the sum in $U_j(z)^{(s)}$($U_j(z)^{(d)}$) is running over the relevant single (double) poles $\alpha_k$ 
and the number of such poles $n_s$($n_d$) depends on that of degeneracies of $\alpha_k$. 
$\Psi (z)=\partial_z \ln \Gamma(z)$ is the digamma function which satisfies 
the identity $\Psi (1-\alpha_k)-\Psi(\alpha_k)=\pi {\rm cot}(\pi \alpha_k)$ and 
$\gamma =-\Psi (1)$ is the Euler-Mascheroni constant. 
For the degenerate $\alpha_k$ in the Picard-Fuchs equation~(\ref{eq:PF}), 
the period vectors $U_j(z)^{(d)}$ have logarithmic terms depending on the complex structure modulus which 
causes the infinite order monodromy transformation for the period vector as explained below.

The obtained period vector is not spanned by the integral symplectic basis which is 
convenient in flux compactification to quantize the three-form fluxes. 
Therefore, along with the strategy of Ref.~\cite{Garcia-Etxebarria:2014wla}, we search for the 
transformation matrix from the basis of Meijer $G$-function to the integral symplectic basis. 
Since the period vector around the LCS point is proportional to the masses of BPS saturated $D_{2p}$-branes 
with $p=0,1,2,3$ corresponding to those of three-brane in the mirror IIB string theory~\cite{Ceresole:1995jg}, 
we can find the integral basis such that the spectrum between 
these D-branes are integral. The authors of Ref.~\cite{Garcia-Etxebarria:2014wla} showed that the integral sympletic basis in the 
large volume limit $z\rightarrow 0$ becomes
\begin{align}
\Pi^{\rm LCS}=\Xi \cdot \pi(z),
\label{eq:Pi}
\end{align}
where $U_j(z)$ is the element of period vector~(\ref{eq:Uj}) around the LCS 
point\footnote{The contour $C$ is chosen to extend from $-i\infty$ to $i\infty$ so as to 
enclose the poles $s=n$.}
\begin{align}
\Xi=
\begin{pmatrix}
1 & 0 & 0 & 0\\
-1 & -1 & 0 & 0\\
-\left( \frac{\kappa b}{4} +\frac{\kappa }{6}\right) & -\left( \frac{\kappa b}{4} +\frac{7\kappa }{6}\right) 
&-2\kappa & -\kappa
\\
-\kappa & -2\kappa & -\kappa & 0
\end{pmatrix}
,
\end{align}
with $b=\frac{1}{3\kappa}\int_{\cal M} c_2 \wedge D$.

In the following, we assume that the above transformation matrix $\Xi$ is applicable for the period vector 
around the SCS point $z\rightarrow \infty$ which is confirmed by considering the 
monodromy transformation of period vector in our discussed concrete models. 
We find the general form of the period vector on the integral symplectic basis near the SCS point 
for the case with degeneracies of $\alpha_i$ less than $2$,
\begin{align}
\Pi^{\rm SCS}&=\Pi^s +\Pi^d,
\label{eq:SCSgeneral}
\end{align}
where 
\begin{align}
\Pi^s &=\sum_{k}^{n_s} \frac{A_k}{\pi^3}(hz)^{-\alpha_k}
\left(
\begin{array}{c}
s_k^3 e^{-\pi i \alpha_k}\\
\frac{i}{2}s_k^2\\
i\frac{\kappa}{8}\left( b s_k^2-(\frac{1}{3}s_k^2 +c_k^2) \right)\\
\frac{s_k\kappa}{4}e^{\pi i \alpha_k}\\
\end{array}
\right)
,\nonumber\\
\Pi^d &=\sum_{k}^{n_d}\frac{A_k}{\pi^3}(hz)^{-\alpha_k}
\left(
\begin{array}{c}
s_k^3 (B_k^0 +\ln h+\ln z -3\pi \frac{c_k}{s_k} +i\pi) e^{-\pi i \alpha_k}\\
\frac{i}{2}s_k^2 (B_k^0+\ln h+\ln z -2\pi \frac{c_k}{s_k})\\
i\frac{\kappa}{8}\left( b s_k^2-(\frac{1}{3}s_k^2 +c_k^2) \right) 
(B_k^0+\ln h+\ln z -2\pi \frac{c_k}{s_k}) -\frac{i\pi c_k \kappa}{4s_k}\\
\frac{s_k\kappa}{4}(B_k^0+\ln h+\ln z -\pi \frac{c_k}{s_k} -i\pi)e^{\pi i \alpha_k}\\
\end{array}
\right)
,
\label{eq:SCS1}
\end{align}
with $s_k={\rm sin}(\pi \alpha_k)$, $c_k={\rm cos}(\pi \alpha_k)$ 
and $A_k=\frac{\Gamma(\alpha_k)^4 \Pi_{i=1,i\neq k,\alpha_i\neq\alpha_k}^4\Gamma (\alpha_i-\alpha_k)}{\Gamma (\alpha_1)\Gamma (\alpha_2)\Gamma (\alpha_3)\Gamma (\alpha_4)}$. 
Here, $\Pi^s$ ($\Pi^d$) encodes the contribution from the single (double) 
poles in the integral of Meijer $G$-function~(\ref{eq:Uj}).


\subsubsection{$\alpha_1\neq \alpha_2 \neq \alpha_3\neq \alpha_4$}
\label{subsubsec:orb}
First of all, we take a closer look at the orbifold point where all $\alpha_i$ are distinct. 
From the formula in Eq.~(\ref{eq:SCS1}) with $n_s=4$ and $n_d=0$, 
the explicit form of period vector becomes 
\begin{align}
\Pi^{\rm SCS}&=\sum_{k=1}^4 \Pi^{s,k} 
=\sum_{k=1}^4 \frac{A_k}{\pi^3}(hz)^{-\alpha_k}
\left(
\begin{array}{c}
s_k^3 e^{-\pi i \alpha_k}\\
\frac{i}{2}s_k^2\\
i\frac{\kappa}{8}\left( b s_k^2-(\frac{1}{3}s_k^2 +c_k^2) \right)\\
\frac{s\kappa}{4}e^{\pi i \alpha_k}\\
\end{array}
\right).
\end{align}

By plugging the above period vector into Eq.~(\ref{eq:Kahler}), 
it enables us to analyze the behavior of the K\"ahler potential around the SCS point
\begin{align}
e^{-K_{\rm SCS}}=\sum_{k=1}^4\frac{A_k^2}{4\pi^6}|hz|^{-2\alpha_k}s_k^3c_k\kappa
\left[ -b s_k^2 +2\left( \frac{2}{3}s_k^2 +c_k^2\right)\right],
\label{eq:orbK}
\end{align}
where the cross terms in the K\"ahler potential such as
\begin{align}
i\Pi^{\dagger s,i}\cdot \Sigma \cdot \Pi^{s,j}
=\frac{A_iA_j}{\pi^6}(hz)^{-\alpha_i}(h\bar{z})^{-\alpha_j}\frac{\kappa}{8}
\biggl[ {\rm sin}((\alpha_i+\alpha_j)\pi)+i(s_i^2-s_j^2)\biggl]
\biggl[(s_i^2+s_j^2)-\left(b+\frac{2}{3}\right)s_i^2s_j^2\biggl]
\end{align}
with $i\neq j$, 
are absent in our all concrete models in Tab.~\ref{tab1}.\footnote{Although the part of obtained K\"ahler potentials is 
different from those of previous results in Ref.~\cite{Candelas:1990rm,Klemm:1992tx} up to overall factors, 
they are caused by the different normalization of complex structure modulus. 
Thus, the K\"ahler metrics are consistent with them when the complex structure modulus is 
redefined as $\alpha \simeq (hz)^{-\alpha_1}$ in the notation of 
Ref.~\cite{Klemm:1992tx}.} 
It then implies that the axionic shift symmetry under 
${\rm arg}(z)\rightarrow {\rm arg}(z) +{\rm const.}$ can be seen in the vicinity of $z\sim \infty$, 
whose property is originating from the invariance under the following monodromy transformation 
around the SCS point:
\begin{align}
\Pi_i^{\rm SCS}(ze^{2\pi i})=\sum_{j=1}^4T_{ij}\Pi_j^{\rm SCS}(z)
\end{align}
with
\begin{align}
T=e^{-2\pi i/d_c}
\begin{pmatrix}
1 & 0 & 0 & 0\\
0 & 1 & 0 & 0\\
0 & 0 & 1 & 0\\
0 & 0 & 0 & 1
\end{pmatrix}
.
\end{align}
Here, $d_c$ is the least common multiple of the degree of homogeneous polynomials corresponding 
to the defining equation of CY manifold. 
We thus find that the monodromy matrix is of finite order,
\begin{align}
T^{d_c}={\bm 1},
\end{align}
as mentioned before.

Following this line of thoughts, we can derive the decay constant 
of closed string axion $\theta={\rm arg}(z)$,
\begin{align}
K_{\theta\theta}=K_{z\bar{z}}|z|^2=
-\frac{A_2^2s_2^3c_2\left(-bs_2^2+2(\frac{2}{3}s_2^2+c_2^2)\right)}{A_1^2s_1^3c_1\left(-bs_1^2+2(\frac{2}{3}s_1^2+c_1^2)\right)}(\alpha_1-\alpha_2)^2h^{2(\alpha_1-\alpha_2)-2}|z|^{2(\alpha_1-\alpha_2)} +{\cal O}(|z|^{2(\alpha_1-\alpha_3)}),
\end{align}
which is valid in the vicinity of $z \sim \infty$. 
It is remarkable that the factors $|z|^2$ in the axion K\"ahler metric appear as a 
fact that the axion is now defined in the phase direction of $z$. 
Since, in all cases, the axion K\"ahler metric vanishes at the SCS point due to the 
inequalities $\alpha_1<\alpha_2<\alpha_3<\alpha_4$, it 
enables us to obtain the small axion decay constant around the SCS point. 
As an example, from the numerical values of K\"ahler metric in Tab.~\ref{tab2}, 
the small axion K\"ahler metric $\sqrt{2K_{\theta\theta}} \simeq 10^{12}\,{\rm GeV}$ 
is achieved under $|z| \simeq 3.7\times 10^{3}$ for the CY manifold 
defined in $\mathbb{P}^4_{2,1,1,1,1}[6]$.
\begin{table}[htb]
\begin{center}
  \begin{tabular}{|c|c|c|} \hline
    CY hypersurfaces & $(\alpha_1, \alpha_2, \alpha_3, \alpha_4)$ & $K_{\theta\theta}$ 
    \\ \hline \hline
     $\mathbb{P}^4_{1,1,1,1,1}[5]$ 
     & $(1/5,2/5,3/5,4/5)$ 
     & $3.1\times 10^{-11}|z|^{-2/5}$\\
     $\mathbb{P}^4_{2,1,1,1,1}[6]$ 
      & $(1/6,2/6,4/6,5/6)$ 
     & $1.4\times 10^{-12}|z|^{-1/3}$\\
    $\mathbb{P}^4_{4,1,1,1,1}[8]$ 
     & $(1/8,3/8,5/8,7/8)$  
    & $3\times 10^{-15}|z|^{-1/2}$\\
    $\mathbb{P}^4_{5,2,1,1,1}[10]$ 
     & $(1/10,3/10,7/10,9/10)$ 
     & $1.2\times 10^{-17}|z|^{-2/5}$
    \\ 
     $\mathbb{P}^5_{6,4,1,1,1,1}[2,12]$ 
     & $(1/12,5/12,7/12,11/12)$ 
    & $1.6\times 10^{-22}|z|^{-2/3}$\\ 
     $\mathbb{P}^5_{3,2,2,1,1,1}[4,6]$ 
     & $(1/4,1/3,2/3,3/4)$ 
    & $3.2\times 10^{-10}|z|^{-1/6}$\\     
     $\mathbb{P}^5_{2,1,1,1,1,1}[3,4]$ 
     & $(1/6,1/4,3/4,5/6)$ 
    & $5.1\times 10^{-13}|z|^{-1/6}$\\ 
     \hline
    \end{tabular}
    \end{center}
        \caption{The axion K\"ahler metric for the several CY threefolds, where all $\alpha_i$ are distinct.} 
    \label{tab2}
\end{table}



\subsubsection{$\alpha_1 \neq \alpha_2 \neq \alpha_3 = \alpha_4$}
\label{subsubsec:hyb1}
In contrast to the previous section, 
we begin with the case where the single pair of $\alpha_i$ is degenerated. 
From the formula in Eq.~(\ref{eq:SCS1}) with $n_s=2$ and $n_d=1$, 
the period vector is explicitly written by
\begin{align}
\Pi^{\rm SCS}&=\sum_{k=1}^2\Pi^{s,k} +\Pi^d,
\end{align}
where 
\begin{align}
\Pi^{s,k} &=\frac{A_k}{\pi^3}(hz)^{-\alpha_k}
\left(
\begin{array}{c}
s_k^3 e^{-\pi i \alpha_k}\\
\frac{i}{2}s_k^2\\
i\frac{\kappa}{8}\left( b s_k^2-(\frac{1}{3}s_k^2 +c_k^2) \right)\\
\frac{s_k\kappa}{4}e^{\pi i \alpha_k}\\
\end{array}
\right)
,\nonumber\\
\Pi^d &=\frac{A_3}{\pi^3}(hz)^{-\alpha_3}
\left(
\begin{array}{c}
s_3^3 (B_3^0 +\ln h+\ln z -3\pi \frac{c_3}{s_3} +i\pi) e^{-\pi i \alpha_3}\\
\frac{i}{2}s_3^2 (B_3^0+\ln h+\ln z -2\pi \frac{c_3}{s_3})\\
i\frac{\kappa}{8}\left( b s_3^2-(\frac{1}{3}s_3^2 +c_3^2) \right) 
(B_3^0+\ln h+\ln z -2\pi \frac{c_3}{s_3}) -\frac{i\pi c_3 \kappa}{4s_3}\\
\frac{s_3\kappa}{4}(B_3^0+\ln h+\ln z -\pi \frac{c_3}{s_3} -i\pi)e^{\pi i \alpha_3}\\
\end{array}
\right)
.
\end{align}

The above period vector leads to the K\"ahler potential for each CY manifold 
by the use of Eq.~(\ref{eq:Kahler}), 
\begin{align}
e^{-K_{\rm SCS}}=\sum_{k=1}^2\frac{A_k^2}{4\pi^6}|hz|^{-2\alpha_k}s_k^3c_k\kappa
\left[ -b s_k^2 +2\left( \frac{2}{3}s_k^2 +c_k^2\right)\right] 
-\frac{\kappa A_3^2}{8\pi^5}|hz|^{-2\alpha_3} \left(b-\frac{4}{3}\right) 
(2B_3 +\ln |hz|^2),
\end{align}
around the SCS point. Here, the cross terms such as $i(\Pi^{\dagger s})^\dagger \cdot \Sigma \cdot \Pi^d$ 
are absent, since they are proportional to the following equalities satisfied in our concrete models in Tab.~\ref{tab1},
\begin{align}
s_k^2\left(b-\frac{1}{3}\right) -1=0,
\end{align}
with $k=1,2$. 
Thus, the obtained K\"ahler potential has the axionic shift symmetry under 
$\theta \rightarrow \theta +{\rm const.}$ with $\theta ={\rm arg}(z)$ which 
is originating from the following fact. 
Around the SCS point, the period vector transforms as
\begin{align}
\Pi^{\rm SCS}(ze^{2\pi i})=\sum_{k=1}^2e^{-2\pi i \alpha_k}\Pi^{s,k} +e^{-2\pi i \alpha_3}\Pi^d 
+2\pi i \xi^{{\rm SCS}},
\end{align}
where 
\begin{align}
\xi^{\rm SCS} &=\frac{A_3}{\pi^3}(hz)^{-\alpha_3}
\left(
\begin{array}{c}
s_3^3 e^{-\pi i \alpha_3}\\
\frac{i}{2}s_3^2\\
i\frac{\kappa}{8}\left( b s_3^2-(\frac{1}{3}s_3^2 +c_3^2) \right)\\
\frac{s_k\kappa}{4}e^{\pi i \alpha_3}\\
\end{array}
\right).
\end{align}

In turn, we find the monodromy matirix $T$ obeying\footnote{In Ref.~\cite{Garcia-Etxebarria:2014wla}, 
the monodromy matrix is also constructed on the Jordan basis of Meijer $G$-function.}
\begin{align}
\Pi_i^{\rm SCS}(ze^{2\pi i})=\sum_{j=1}^4T_{ij}\Pi_j^{\rm SCS}(z),
\end{align}
generically satisfies the following equality, even if we do not know its explicit form\footnote{The explicit form of monodromy matrix $T$ can be derived as follows. First, we obtain the 
monodromy matrix of the period vector around the SCS point on the basis of Eq.~(\ref{eq:piJ}), 
known as the {\it Jordan basis}~\cite{Greene:2000ci}. 
Then, by acting the transformation matrix $\Xi$ in Eq.~(\ref{eq:Pi}) 
on the obtained monodromy matrix of the Jordan basis, the monodromy matirix $T$ 
is obtained.}
\begin{align}
(T^{d_c}-{\bm 1})^2=0,
\end{align}
which corresponds to the symmetry of K\"ahler potential for each CY manifold. 
The above equality is originating from the fact that the operator $-(T^{d_c})^2+2T^{d_c}$ is identified 
with the identity operator, i.e., ${\bm 1}$. 

The K\"ahler metric of axion is then obtained as
\begin{align}
K_{\theta\theta}&=K_{z\bar{z}}|z|^2\nonumber\\
&\simeq 
\frac{\pi A_3^2\left(b-\frac{4}{3}\right)}{2A_1^2s_1^3c_1\left(-bs_1^2+2(\frac{2}{3}s_1^2+c_1^2)\right)}
\biggl[ (\alpha_1-\alpha_3)^2(2B_3^0 +\ln |hz|^2) +\alpha_1-\alpha_3+1\biggl]|hz|^{2(\alpha_1-\alpha_3)},
\end{align}
which is valid in the vicinity of $z \sim \infty$. 
Since the axion K\"ahler metric vanishes at the SCS point of 
CY manifold due to the inequalities  $\alpha_1<\alpha_3=\alpha_4<\alpha_2$ as shown in Tab.~\ref{tab1}, 
we can obtain the small axion decay constant around the SCS point. 
In Tab.~\ref{tab3}, we show the numerical values of K\"ahler metric for 
the CY manifold in Tab.~\ref{tab1}. 
As an example, from the numerical values of K\"ahler metric in Tab.~\ref{tab3}, 
the small axion K\"ahler metric $\sqrt{2K_{\theta\theta}} \simeq 10^{12}\,{\rm GeV}$ 
is achieved under $|z| \simeq 7.2\times 10^{13}$ for the CY manifold 
defined in $\mathbb{P}^5_{3,1,1,1,1,1}[2,6]$.
\begin{table}[htb]
\begin{center}
  \begin{tabular}{|c|c|c|} \hline
    CY hypersurfaces & $(\alpha_1, \alpha_2, \alpha_3, \alpha_4)$ & $K_{\theta\theta}$ 
    \\ \hline \hline
    $\mathbb{P}^5_{1,1,1,1,1,1}[2,4]$ & $(1/4,3/4,1/2,1/2)$ 
     & $9.3\times 10^{-5}|z|^{-1/2}\ln|z|$\\
     $\mathbb{P}^6_{1,1,1,1,1,1,1}[2,2,3]$ & $(1/3,2/3,1/2,1/2)$ 
     & $3.6\times 10^{-4}|z|^{-1/3}\ln|z|$\\
    $\mathbb{P}^5_{3,1,1,1,1,1}[2,6]$ & $(1/6,5/6,1/2,1/2)$ 
    & $4.7\times 10^{-6}|z|^{-2/3}\ln|z|$\\
     \hline
    \end{tabular}
    \end{center}
        \caption{The axion K\"ahler metric for the several CY threefolds, where one pair of $\alpha_i$ 
        is degenerated.} 
    \label{tab3}
\end{table}

\subsubsection{$\alpha_1 = \alpha_2 \neq \alpha_3 = \alpha_4$}
\label{subsubsec:hyb2}
Next, we consider the CY manifold with $\alpha_1 = \alpha_2 \neq \alpha_3 = \alpha_4$ 
which causes the double poles around the SCS point. 

Following the same procedure in Sec.~\ref{subsubsec:hyb1}, 
the period vector on the integral symplectic basis is obtained from the formula in 
Eq.~(\ref{eq:SCS1}) with $n_s=0$ and $n_d=2$, 
\begin{align}
\Pi^{\rm SCS}=
\sum_{k=1,3}\Pi^{d,k} =\sum_{k=1,3} \frac{A_k}{\pi^3}(hz)^{-\alpha_k}
\left(
\begin{array}{c}
s_k^3 (B_k^0 +\ln h+\ln z -3\pi \frac{c_k}{s_k} +i\pi) e^{-\pi i \alpha_k}\\
\frac{i}{2}s_k^2 (B_k^0+\ln h+\ln z -2\pi \frac{c_k}{s_k})\\
i\frac{\kappa}{8}\left( b s_k^2-(\frac{1}{3}s_k^2 +c_k^2) \right) 
(B_k^0+\ln h+\ln z -2\pi \frac{c_k}{s_k}) -\frac{i\pi c_k \kappa}{4s_k}\\
\frac{s_k\kappa}{4}(B_k^0+\ln h+\ln z -\pi \frac{c_k}{s_k} -i\pi)e^{\pi i \alpha_k}\\
\end{array}
\right)
,
\end{align}
which leads to the K\"ahler potential
\begin{align}
e^{-K_{\rm SCS}}=
\sum_{k=1,3}\frac{\kappa A_k^2}{2\pi^5}|hz|^{-2\alpha_k}s_k^2c_k^2
\left[ \ln |hz|^2 +2B_k -6\pi \frac{c_k}{s_k} +\frac{\pi}{s_kc_k}\right].
\end{align}
Here, the cross term $i\Pi^{\dagger d,1} \cdot \Sigma \cdot \Pi^{d,3}$ 
is absent by the use of following equalities satisfied in our concrete models in Tab.~\ref{tab1}, 
\begin{align}
s_k^2\left(b+\frac{2}{3}\right) -2=0,
\end{align}
with $k=1,3$. 
Thus, the obtained K\"ahler potential has the axionic shift symmetry under 
$\theta \rightarrow \theta +{\rm const.}$ with $\theta ={\rm arg}(z)$ which 
is originating from the following fact.

Around the SCS point, the period vector transforms as
\begin{align}
\Pi^{\rm SCS}(ze^{2\pi i})=\sum_{k=1,3}e^{-2\pi i \alpha_k}\Pi^{d,k} +2\pi i \sum_{k=1,3}\xi^{s,k},
\end{align}
where 
\begin{align}
\xi^{s,k} &=\frac{A_k}{\pi^3}(hz)^{-\alpha_k}
\left(
\begin{array}{c}
s_k^3 e^{-\pi i \alpha_k}\\
\frac{i}{2}s_k^2\\
i\frac{\kappa}{8}\left( b s_k^2-(\frac{1}{3}s_k^2 +c_k^2) \right)\\
\frac{s_k\kappa}{4}e^{\pi i \alpha_k}\\
\end{array}
\right).
\end{align}

In turn, we find the monodromy matirix $T$ obeying
\begin{align}
\Pi_i^{\rm SCS}(ze^{2\pi i})=\sum_{j=1}^4T_{ij}\Pi_j^{\rm SCS}(z),
\end{align}
generically satisfies the following equality, even if we do not know its explicit form,
\begin{align}
(T^{d_c}-{\bm 1})^2=0,
\end{align}
which corresponds to the symmetry of K\"ahler potential for each CY manifold 
in the same way with Sec.~\ref{subsubsec:hyb1}. 
The K\"ahler metric of axion is then obtained in the vicinity of $z \sim \infty$,
\begin{align}
K_{\theta\theta}=K_{z\bar{z}}|z|^2&\simeq 
(\alpha_1)^2\left(\frac{\left[ \ln |hz|^2 +2B_1 -6\pi \frac{c_1}{s_1} +\frac{\pi}{s_1c_1}-\frac{1}{\alpha_1}\right]^2}
{\left[ \ln |hz|^2 +2B_1 -6\pi \frac{c_1}{s_1} +\frac{\pi}{s_1c_1}\right]^2}
-
\frac{\ln |hz|^2 +2B_1 -6\pi \frac{c_1}{s_1} +\frac{\pi}{s_1c_1}-\frac{2}{\alpha_1}}
{\ln |hz|^2 +2B_1 -6\pi \frac{c_1}{s_1} +\frac{\pi}{s_1c_1}}\right)
\nonumber\\
&+{\cal O}(|z|^{2(\alpha_1-\alpha_3)}\ln|z|^2),
\end{align}
from which the axion K\"ahler metric vanishes at the SCS point of 
CY manifold by using the inequalities  $\alpha_1=\alpha_2<\alpha_3=\alpha_4$. 
In contrast to the previous case, constrained axion K\"ahler metric $\sqrt{2K_{\theta\theta}}\simeq {\cal O}(10^{12})\,{\rm GeV}$ requires 
the value of complex structure modulus $|z|$ is of ${\cal O}(10^{300})$ due to the 
logarithmic dependence of the K\"ahler metric. 

\subsubsection{$\alpha_1 = \alpha_2 = \alpha_3 = \alpha_4$}
\label{subsubsec:hyb3}
In this section, we consider the CY manifold with $\alpha_1 = \alpha_2 = \alpha_3 = \alpha_4$ 
which causes the quadrupole poles around the SCS point as shown in the Meijer integral~(\ref{eq:Uj}). 
So far, the SCS point can be distinguishable from the LCS point in the CY manifold. 
However, in the current case, Picard-Fuchs equation around the SCS point is 
related to that around the LCS point. 
Along the line of Ref.~\cite{Lazaroiu:2000jx}, 
under the following change of variable $z$ and solution of Picard-Fuchs equation in Eq.~(\ref{eq:PF}):
\begin{align}
u&= z^{-1},
\nonumber\\
\tilde{\pi} (u)&=z^{-\alpha_1}\pi (z),
\end{align}
we have obtain the same Picard-Fuchs equation~(\ref{eq:PF}),
\begin{align}
\left\{\delta_u^4
-hu(\delta_u +\alpha_1)^4\right\}
\pi (u)=0,
\label{eq:PFu}
\end{align}
where $\delta_u=ud/du=-zd/dz$. 
It implies that the SCS point is physically equivalent to the LCS point. 
Indeed, when the contour $C$ in Eq.~(\ref{eq:Uj}) is taken to enclose such quadrupole poles, 
the Meijer $G$-function has the logarithmic terms~\cite{Lazaroiu:2000jx}:
\begin{equation}
U_{j}(z)=\frac{1}{(2 \pi i)^j} \left( \frac{\sin \pi \alpha}{\pi} \right)^{3-j}(e^{\pi i(j+1)}hz)^{-\alpha_1} 
\left[ (\log (hz))^3+D_{2}^{j} (\log (hz))^2+D_{1}^{j} \log (hz)+D_{0}^{j} \right],
\end{equation}
where
\begin{align}
D_{0}^{j}&=\frac{32}{3}\gamma^3+\frac{4}{3}\gamma \pi^2 +\frac{4}{3} \xi(2)+(8\gamma^2+\frac{1}{3}\pi^2)C^{j}+4\gamma(C^{j2}+C^{j'})+C^{j3}+3C^{j}C^{j'}+C^{j''},
\nonumber\\
D_{1}^{j}&=8\gamma^2+\frac{1}{3}\pi^2+8\gamma C^{j}+3(C^{j})^3+3C^{j'},
\nonumber\\
D_{2}^{j}&=4\gamma+3C^{j},
\end{align}
with 
\begin{align}
C^j&\equiv {\cal C}^{j}\bigl|_{s=-\alpha},
\nonumber\\
C^{j'} &\equiv \frac{d}{ds}{\cal C}^{j}\biggl|_{s=-\alpha},
\nonumber\\
C^{j''} &\equiv \frac{d^2}{ds^2}{\cal C}^{j}\biggl|_{s=-\alpha},
\nonumber\\
{\cal C}^{j}&=\log k+(j+1)i\pi -(j+1)\Psi(-s)-(j-3)\Psi(1+s).
\end{align}

Such cubic term of logarithm induced by the existence of quadrupole poles, 
also appears in the solution of Picard-Fuchs equation around the LCS point 
as discussed in Sec.~\ref{subsec:LCS}. 
Thus, even if the small K\"ahler metric is obtained around the SCS point, 
the K\"ahler metric corresponds to that around the large complex modulus limit. 
We analyze the detail of such a case in Sec.~\ref{subsec:LCS}. 


\subsection{Conifold point}
\label{subsec:coni}
In a way similar to the previous sections, 
we next analyze the decay constant of closed string axion around other special point, in particular, conifold point 
which often appears in the landscape of string theory~\cite{Hebecker:2006bn} and the vicinity of large number of D$3$-branes~\cite{Verlinde:1999fy}. 
For our purpose, we study the type IIB string theory compactified on the mirror quintic CY threefold which 
encodes the conifold point. However, the solution of Picard-Fuchs equation around the conifold point 
cannot be described by the Meijer $G$-function. 

To obtain the period vector around the conifold point, we directly solve the Picard-Fuchs equation 
around the conifold point. First of all, when we redefine the complex structure modulus 
as $z_{\rm c}=1-\psi^{-5}$ in the defining equation of mirror quintic, 
\begin{align}
{\cal P}_1&=(x_0)^5+(x_1)^5+(x_2)^5+(x_3)^5+(x_4)^5-5 \psi \Pi_{i=0}^4 x_i=0,~~~~(x_i\in \mathbb{P}^4_{1,1,1,1,1}[5]),
\label{eq:P1}
\end{align}
associated Picard-Fuchs equation in Eq.~(\ref{eq:PF}) reduces to
\begin{align}
\left\{  P_4(z_{\rm c})\theta_{\rm c}^4 +P_3(z_{\rm c})\theta_{\rm c}^3 +P_2(z_{\rm c})\theta_{\rm c}^2 
+P_1(z_{\rm c})\theta_{\rm c}^1+P_0(z_{\rm c})\right\} \pi^c(z_{\rm c})=0,
\end{align}
where $\theta_{\rm c} =z_{\rm c}d/dz_{\rm c}$ and 
\begin{align}
P_4(z_{\rm c})&=1-3z_{\rm c} +3z_{\rm c}^2-z_{\rm c}^3,
\nonumber\\
P_3(z_{\rm c})&=-4+6z_{\rm c} -14z_{\rm c}^3,
\nonumber\\
P_2(z_{\rm c})&=\frac{1}{5}\left(25-22z_{\rm c} +4z_{\rm c}^2+103z_{\rm c}^3\right),
\nonumber\\
P_1(z_{\rm c})&=\frac{1}{5}\left(-10+7z_{\rm c} -2z_{\rm c}^3\right),
\nonumber\\
P_0(z_{\rm c})&=-\frac{24}{5^5}z_{\rm c}^3.
\end{align}
When we take the following ansatz of solution of Picard-Fuchs equation, 
\begin{align}
\pi^{\rm c}(z_{\rm c})=\sum_{n=0}^\infty \alpha_n z_{\rm c}^{\rho +n},
\end{align}
the characteristic exponent of Picard-Fuchs operator is obtained as $\rho=0,1,1,2$ 
which results in the period vector by using the recursive approach,
\begin{align}
\pi^{\rm c}(z_{\rm c})=
\begin{pmatrix}
\pi^{\rm c}_1
\\
\pi^{\rm c}_2
\\
\pi^{\rm c}_3
\\
\pi^{\rm c}_4
\end{pmatrix}
=
\begin{pmatrix}
1+\frac{2}{5^4}z_{\rm c}^3 +\frac{97}{2\cdot 4\cdot 5}z_{\rm c}^4+\cdots
\\
z_{\rm c}+\frac{7}{10}z_{\rm c}^2 +\frac{41}{75}z_{\rm c}^3 +\frac{1133}{4\cdot 5^5}z_{\rm c}^4+\cdots
\\
z_{\rm c}^2+\frac{37}{30}z_{\rm c}^3 +\frac{2309}{1800}z_{\rm c}^4+\cdots
\\
\pi_2 \ln z -\frac{23}{360}z_{\rm c}^3 -\frac{6397}{3\cdot 10^6}z_{\rm c}^4+\cdots
\end{pmatrix}
,
\label{eq:periodc}
\end{align}
However, the obtained period vector is not spanned by the integral symplectic basis. 
By acting the transformation matrix\footnote{The elements of transformation 
matrix is only numerically known in Ref.~\cite{Huang:2006hq} as $a_1=6.19501627714957$, $a_2= 1.016604716702582$, 
$a_3= −0.140889979448831$, $a_4=1.07072586843016$, $a_5=-0.0247076138044847$,
$a_6= 1.29357398450411$, $a_7=\frac{2a_2a_6\pi -\sqrt{5}a_4}{2a_1\pi}$, 
$a_8=\frac{5+16a_3a_6\pi^3}{16a_1\pi^3}$, $a_9=\frac{\sqrt{5}a_2 +8a_3a_4\pi^2}{8a_1\pi^2}$.} 
\begin{align}
\Xi^{\rm c}=
\begin{pmatrix}
a_4 & a_5 & a_9  &-\frac{\sqrt{5}}{(2\pi i)^2}
\\
ia_6 & ia_7 & ia_8 & 0
\\
0 & \frac{\sqrt{5}}{2\pi i} & 0 & 0
\\
a_1 -\frac{11}{2}i a_6 & a_2-\frac{11}{2}i a_7 & a_3-\frac{11}{2}i a_8 & 0
\end{pmatrix}
,
\end{align}
with the period vector~(\ref{eq:periodc}),
the period vector on the integral symplectic basis can be obtained around the conifold point
\begin{align}
\Pi^{\rm c}=
\begin{pmatrix}
-\frac{\sqrt{5}}{(2\pi i)^2}z_{\rm c}\ln z_{\rm c} +a_4 +a_5z_{\rm c} +{\cal O}(z_{\rm c}^2)
\\
ia_6 +a_{11}z_{\rm c} +{\cal O}(z_{\rm c}^2)
\\
\frac{\sqrt{5}}{2\pi i} z_{\rm c} +{\cal O}(z_{\rm c}^2)
\\
a_{12} +a_{10} z_{\rm c}+{\cal O}(z_{\rm c}^2)
\end{pmatrix}
,
\label{eq:Pic}
\end{align}
with $a_{10}=1.016604716702582-0.8292168231795108i$ and 
$a_{11}=0.15076669512354743i$, and $a_{12}=a_1-7.114656914772605i$. 

From the K\"ahler potential
\begin{align}
e^{-K}=i (\Pi^{\rm c})^\dagger\cdot \Sigma \cdot \Pi^{\rm c}
=\frac{5}{(2\pi)^3}|z_{\rm c}|^2\ln|z_{\rm c}|^2 +C+D(z_{\rm c} +\bar{z}_{\rm c})
+E|z_{\rm c}|^2
,
\end{align}
with $C\simeq 16.02$, $D\simeq 2.63$ and $E\simeq 0.289$, we identify 
the phase of complex structure modulus as axion $z_{\rm c}=e^{2\pi i\theta_{\rm c}}$, 
since the K\"ahler potential has the discrete shift symmetry under 
$\theta_{\rm c} \rightarrow \theta_{\rm c} +N$ with $N$ being integer. 
Such a symmetry is originating from the following monodromy transformation. 
Around the conifold point, the period vector transforms as
\begin{align}
\Pi^{\rm c}(ze^{2\pi i})=\Pi^{\rm c}(z) +\xi^{\rm c}(z),
\end{align}
where 
\begin{align}
\xi^{\rm c}&=-\frac{5}{(2\pi i)^2}z_{\rm c} 
\left(
\begin{array}{c}
1\\
0\\
0\\
0
\end{array}
\right).
\end{align}
Consequently, under the monodromy transformation:
\begin{align}
\Pi_i^{\rm c}\rightarrow \sum_{j=1}^4T_{ij}^{\rm c}\Pi_j^{\rm c},
\end{align}
we find that the transformation matrix $T^{\rm c}$ satisfies the following equalities, even if we do not know its explicit form,\footnote{The explicit form of $T^{\rm c}$ is obtained in a similar way with Sec.~\ref{subsubsec:hyb1}.}
\begin{align}
(T^{\rm c}-{\bm 1})^2=0,
\end{align}
which corresponds to the symmetry of K\"ahler potential in the same way with 
Secs.~\ref{subsubsec:hyb1} and \ref{subsubsec:hyb2}.

In the following, we proceed to study the decay constant of axion $\theta_{\rm c}$. 
The K\"ahler metric of axion $\theta_{\rm c}$ becomes
\begin{align}
K_{\theta_{\rm c}\theta_{\rm c}}=K_{z_{\rm c}\bar{z}_{\rm c}}|z_{\rm c}|^2=
\frac{|z_{\rm c}|^2}{V_{\rm c}^2}\Biggl|\frac{5}{(2\pi)^3}\left(z_{\rm c}\ln|z_{\rm c}|^2 +z_{\rm c}\right) 
+D+E z_{\rm c}\Biggl|^2 -\frac{|z_{\rm c}|^2}{V_{\rm c}}\left( \frac{5}{(2\pi)^3}\ln|z_{\rm c}|^2 +\frac{10}{(2\pi)^3} 
+E\right)
,
\label{eq:Kmetricconi}
\end{align}
with $V_{\rm c}=e^{-K}$, which approaches to zero in the limit of $z_{\rm c}\rightarrow 0$. 
We will thus expect that the axion decay constant can be taken much smaller than the 
string scale. 
Indeed, from the K\"ahler metric in Eq.~(\ref{eq:Kmetricconi}), 
the small axion K\"ahler metric $\sqrt{2K_{\theta_{\rm c}\theta_{\rm c}}} \simeq 10^{12}\,{\rm GeV}$ 
is achieved under $|z_{\rm c}| \simeq 1.6\times10^{-6}$ for the mirror quintic CY 
defined in $\mathbb{P}^4_{1,1,1,1,1}[5]$.
%


\subsection{LCS point}
\label{subsec:LCS}
Finally, we discuss the decay constant of single and multiple axions around the 
LCS point based on the 
effective action of type IIA string theory on CY manifold $\tilde{{\cal M}}$. 
By applying the mirror map for the period vector of Picard-Fuchs equation around the LCS point,
$t^i\simeq \frac{\ln z^i}{2\pi i}$ with $i=1,2,\cdots, h^{2,1}$, 
we can obtain the period vector around the large volume point in the type IIA string theory. 
The authors of Ref.~\cite{Hosono:1993qy,Hosono:1994ax} showed that the period vector around the large volume 
point is represented by
\begin{align}
\Pi^{\rm LCS}=
\begin{pmatrix}
1\\
t^i\\
2F- t^i\partial_i F\\
\partial_i F
\end{pmatrix}
,
\label{eq:PiLCS}
\end{align}
where $F$ is the instanton-corrected prepotential\footnote{Here and in what follows, 
the summation over $i,j,k=1,2,\cdots, h^{2,1}$ is implied.}
\begin{align}
F=-\frac{1}{3!}\kappa_{ijk}t^it^jt^k -\frac{1}{2}\kappa_{ij}t^it^j +\kappa_i t^i +\frac{1}{2}\kappa_0 
-\frac{1}{(2\pi i)^3}\sum_{\beta}n_\beta {\rm Li}_3(q^{\beta}),
\end{align}
with ${\rm Li}_s(z)=\sum_{n=1}\frac{z^n}{n^s}$ being polylogarithm function. 
Here, $q^{\beta_i}=e^{2\pi id_it^i}$ denote the instanton corrections 
labeled by instanton number $n_\beta$, where $\beta=d_i \beta_i$ 
with $d_i$ being integers is expanded on the integral cohomology basis for 
$H_2(\tilde{{\cal M}},\mathbb{Z})\backslash \{0\}$ of the CY manifold $\tilde{{\cal M}}$. 
Furthermore, the topological invariant quantities are also defined in terms of the K\"ahler 
form $J_i$ of $\tilde{{\cal M}}$,
\begin{align}
&\kappa_{ijk}=\int_{\tilde{{\cal M}}} J_i \wedge J_j \wedge J_k, \qquad 
\kappa_{ij} =-\frac{1}{2}\int_{\tilde{{\cal M}}} J_i \wedge J_j \wedge J_j, \nonumber\\
&\kappa_j =\frac{1}{24}\int_{\tilde{{\cal M}}} c_2({\tilde{{\cal M}}}) \wedge J_j, \qquad 
\kappa_0 =\frac{\zeta(3)}{(2\pi i)^3}\int_{\tilde{{\cal M}}} c_3({\tilde{{\cal M}}}) 
=\frac{\zeta(3)}{(2\pi i)^3}\chi ({\tilde{{\cal M}}}),
\end{align}
where $\zeta(3)\simeq 1.2$ and $\chi ({\tilde{{\cal M}}})$ is the Euler characteristic.

As shown below, the K\"ahler potential constructed by the period vector~(\ref{eq:PiLCS}) 
involves the axionic shift symmetries $t^i\rightarrow t^i+1$, 
originating from the gauge symmetry of Kalb-Ramond field ${\rm Re}\,t^i$. 
Indeed, the period vector (\ref{eq:PiLCS}) around the LCS point is transformed under 
$t^i\rightarrow t^i+1$ for $i=1,2,\cdots, h^{2,1}$,
\begin{align}
\Pi^{\rm LCS}(t^i+1)=T^{\rm LCS}\cdot \Pi^{\rm LCS}(t^i),
\end{align}
where 
\begin{align}
T^{\rm LCS}=
\begin{pmatrix}
{\cal T}_1 & {\bm 0}\\ 
{\cal T}_2 & {\cal T}_3
\end{pmatrix}
,
\end{align}
with 
\begin{align}
{\cal T}_1&=
\begin{pmatrix}
1 & 0 & 0 & \cdots &0 \\ 
1 & 1 & 0  & \cdots &0\\
1 & 0 & 1 & \cdots &0\\
\vdots & \vdots & \vdots &\ddots &\vdots \\
1 & 0 & 0  & \cdots & 1
\end{pmatrix}
,
{\cal T}_2=
\begin{pmatrix}
\sum_{i}\kappa_{i} & {\cal A}_1& \cdots
& {\cal A}_{h^{2,1}}\\
-{\cal A}_1
&-\sum_{j}\kappa_{1j1} & \cdots & -\sum_{j}\kappa_{1j h^{2,1}} \\
\vdots & \vdots & \vdots &\vdots  \\
-{\cal A}_{h^{2,1}}
&-\sum_{j}\kappa_{h^{2,1}j1} & \cdots & -\sum_{j}\kappa_{h^{2,1}j h^{2,1}} 
\end{pmatrix}
,
\nonumber\\
{\cal T}_3&=
\begin{pmatrix}
1 &-1 & -1 & \cdots &-1\\
0 &1 & 0  &\cdots &0 \\ 
0 & 0 & 1 & \cdots &0\\
\vdots & \vdots & \vdots &\ddots & \vdots \\
0 & 0 & 0 &\cdots & 1 
\end{pmatrix}
,
\end{align}
and ${\cal A}_i=\frac{1}{2}\left(\sum_{j,k}\kappa_{ijk}-\sum_{j}(\kappa_{ij}+\kappa_{ji})\right)$.
The above monodromy matrix $T^{\rm LCS}$ 
satisfies the following maximally unipotent property,
\begin{align}
(T^{\rm LCS}-{\bm 1})^{2(h^{2,1}+1)}=0,
\end{align}
which leads to the symmetries of the K\"ahler potential as shown below. 


In the remaining of this section, we analyze the axion decay constant around the LCS point 
with single axion in Sec.~\ref{subsubsec:LCS1} and multiple axions in Sec.~\ref{subsubsec:LCS2}.

\subsubsection{Single axion}
\label{subsubsec:LCS1}
We begin with the single axion case. 
From the period vector in Eq.~(\ref{eq:PiLCS}), 
the quantum-corrected K\"ahler potential for single modulus $T=it$ is represented by
\begin{align}
K(T,\bar{T}) =-\ln &\biggl[ \frac{1}{6}\kappa_{ttt} (T +\bar{T})^3
-\frac{\zeta(3)}{4\pi^3}\chi (\tilde{{\cal M}}) \nonumber\\
&+\frac{2}{(2\pi )^2} \sum_{n,d_t=1}^\infty  \frac{d_t n_\beta}{n^2} (T+\bar{T}) e^{-\pi n d_t (T+\bar{T})} 
{\rm cos}\left(-i\pi n d_t (T-\bar{T})\right)
\nonumber\\
&+\frac{4}{(2\pi)^3} \sum_{n,d_t=1}^\infty  \frac{n_\beta}{n^3} e^{-\pi n d_t (T+\bar{T})} 
{\rm cos}\left(-i\pi n d_t (T-\bar{T})\right)\biggl],
\label{eq:Ks}
\end{align}
in the large volume limit, 
where $\zeta(3)\simeq 1.2$, $\kappa_{ttt}$ denotes the classical intersection number of CY manifold, 
and 
the exponential terms show the instanton corrections labeled by the instanton number $n_\beta$ and 
the degree of internal two-cycle $d_t$. 
Since as mentioned before, there is a discrete axionic shift 
symmetry under ${\rm Im}\,T\rightarrow {\rm Im}\,T +N$ with $N$ being integer, 
we identify the axion as ${\rm Im}\,T$ and its decay constant as $f\simeq \sqrt{2K_{T\bar{T}}}/8\pi^2$. 
The kinetic term including the instanton correction is extracted from the above K\"ahler potential, 
\begin{align}
K_{T\bar{T}} ={\rm Re}(K_T) \left({\rm Re}(K_T) +\frac{2}{T+\bar{T}}\right) +{\rm Im}(K_T)^2,
\end{align}
where
\begin{align}
K_{T}=&-e^{K}\Biggl[\frac{\kappa_{ttt}(T+\bar{T})^2}{2}-\sum_{n,d_t=1}^\infty
\frac{d_t^2n_\beta(T+\bar{T})}{2\pi n}e^{-\pi n d_t (T+\bar{T})} 
{\rm cos}\left(-i\pi n d_t (T-\bar{T})\right)
\nonumber\\
&+i
\sum_{n,d_t=1}^\infty n_\beta \left(
\frac{2d_t^2(T+\bar{T})}{2\pi n} +\frac{2d_t}{(2\pi)^2n^2}\right)e^{-\pi n d_t (T+\bar{T})} 
{\rm sin}\left(-i\pi n d_t (T-\bar{T})\right)\Biggl].
\label{eq:KT1}
\end{align}

\begin{figure}[htbp] 
 \begin{minipage}{0.5\hsize}
  \begin{center}
   \includegraphics[width=80mm]{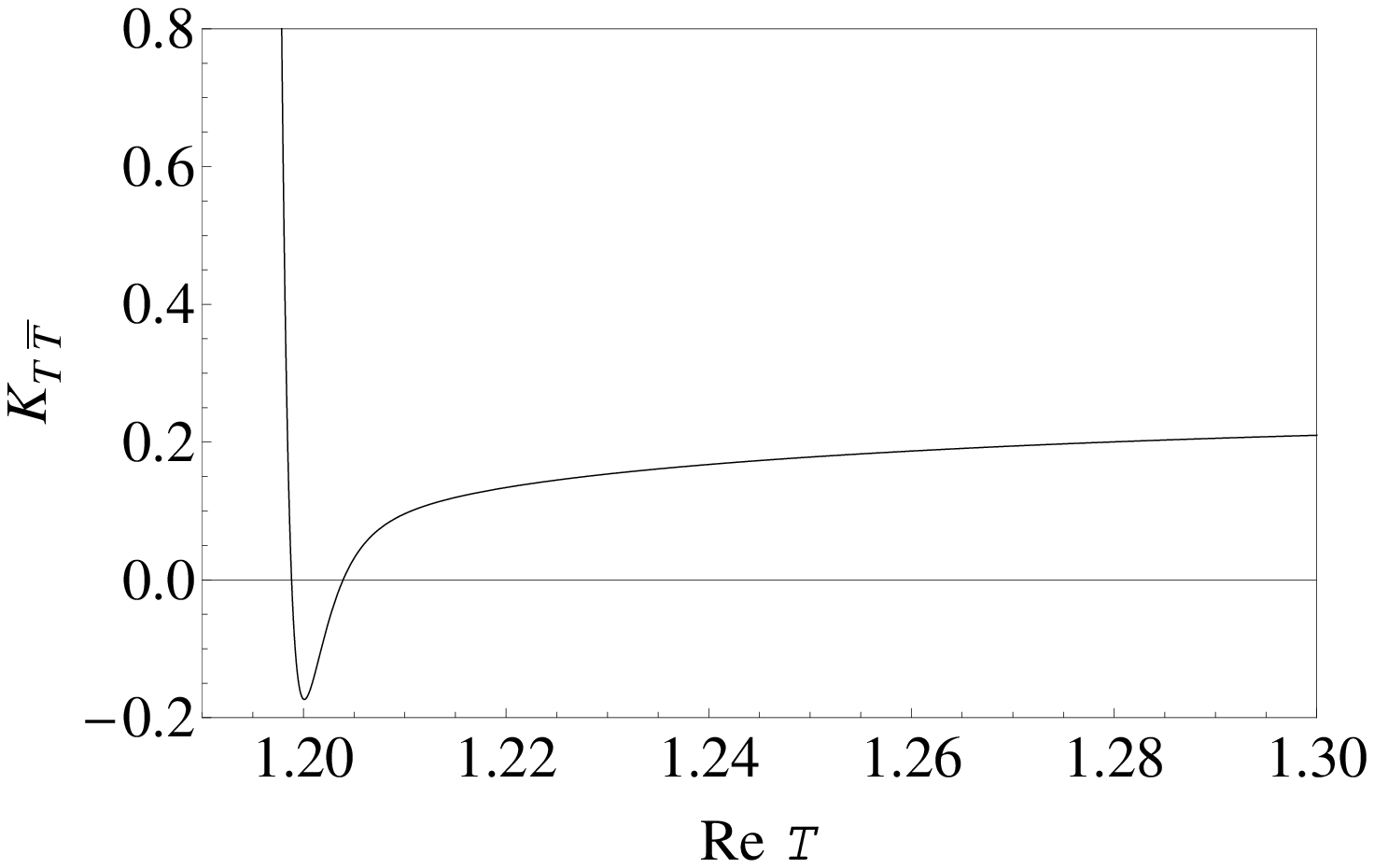}
  \end{center}
  \end{minipage}
 \begin{minipage}{0.5\hsize}
  \begin{center}
   \includegraphics[width=80mm]{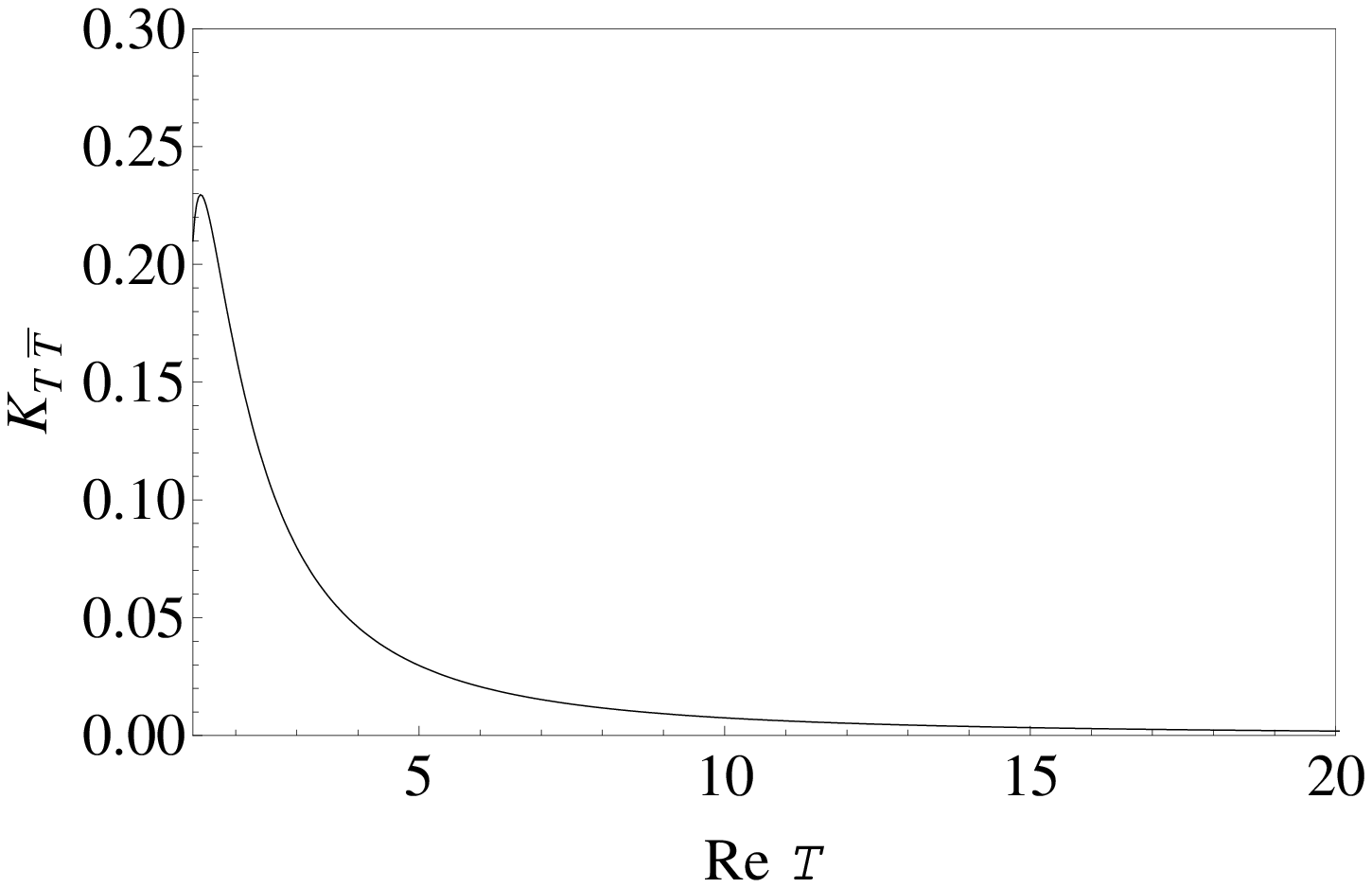}
  \end{center}
  \end{minipage}
    \caption{In both panels, the black curve corresponds to the quantum-corrected K\"ahler metric on the (${\rm Re}\,T, K_{T\bar{T}}$)-plane for the mirror quintic CY manifold~\cite{Candelas:1990rm}. 
    In the left (right) panel, the K\"ahler metric is plotted within the range, $1.18\leq {\rm Re}T \leq 1.3$ 
    ($1.3\leq {\rm Re}T \leq 20$).}
 \label{fig:1}
\end{figure}

In Fig.~\ref{fig:1}, we draw the quantum-corrected K\"ahler metric on the hypersurface ${\rm Im}\,T=0$ 
for single representative CY manifold, i.e., mirror quintic with $\chi =-200$~\cite{Candelas:1990rm}. 
Now, we take into account the instanton effects up to order one hundred for 
mirror quintic by using the numerical code~\cite{Klemm}. 
It is then found that the moduli K\"ahler metric has vanished at two points 
around the moduli value ${\cal O}(1)$ in the left panel in Fig.~\ref{fig:1}, 
where the smaller modulus value leading to the vanishing K\"ahler metric satisfies 
\begin{align}
K_{T}&=0,
\label{eq:cons1}
\end{align}
whereas the larger one satisfies 
\begin{align}
{\rm Re}(K_{T})+\frac{2}{T+\bar{T}}&=0,
\nonumber\\
{\rm Im}\,T &=0.
\label{eq:cons2}
\end{align}
In this way, although the vanishing K\"ahler metric is unphysical, 
the tiny moduli K\"ahler metric is achieved at the values close to those of Eqs.~(\ref{eq:cons1}) 
or (\ref{eq:cons2}). 
It is then expected that the small axion decay constant is obtained around these points. 
Recall that, the negative instanton contributions in Eq.~(\ref{eq:KT1}) 
give rise to the small value of $K_T$ in general class of CY manifolds.

However, we have to take into account the fundamental domain where 
the large volume limit is applicable. 
To make the analysis concrete, we focus on the mirror quintic CY defined in Eq.~(\ref{eq:P1}), where 
there are three special points such as orbifold, conifold, 
and LCS points, corresponding to $\psi=0,1,\infty$. 
The complex structure modulus $\psi$ in B-model is in one-to-one 
correspondence with K\"ahler modulus $T$ in the A-model under the mirror map~\cite{Candelas:1990rm}:
\begin{align}
T(\psi)=-\frac{5}{2\pi}\left( 
\ln (5\psi)-\frac{1}{w_0^{(1)}}\sum_{m=1}^\infty \frac{(5m)!}{(m!)^5 (5\psi)^{5m}}
\bigl[ \Psi (1+5m) -\Psi (1+m)\bigl] \right),
\end{align}
with $\Psi$ being the digamma function, 
and accordingly the quantum-corrected Yukawa coupling $\hat{\kappa}_{ttt}$ behaves in the limit $\psi \rightarrow 1$,
\begin{align}
\hat{\kappa}_{ttt} &=5+\sum_{d_t=1}^\infty \frac{n_{d_t} d_t^3 e^{-2\pi d_tT}}{1-e^{-2\pi  d_tT}}
\nonumber\\
&\propto \frac{1}{(\psi-1)\bigl[-\ln (\psi-1)\bigl]^3}
\end{align}
from which the Yukawa coupling diverges at the conifold point. 
Thus, the convergence radius in the large volume limit is determined so 
as to avoid such a singularity in the Yukawa coupling~\cite{Candelas:1990rm}, i.e., 
\begin{align}
{\rm Re}\,T > {\rm Re}\,T(1)\simeq 1.208.
\end{align}

From this considerations, we conclude that the two points where the K\"ahler metric 
vanishes in the left panel in Fig.~\ref{fig:1} are not in the large volume region, but in the conifold region. 
Although we find that these vanishing K\"ahler metric around the above 
similar points can be seen for the other one-parameter 
CY manifolds, the large volume expansion is not valid. 
Thus, the tiny axion decay constant is only obtained around the large volume point 
${\rm Re}\,T\rightarrow \infty$ leading to $K_{T\bar{T}}\rightarrow 0$ as shown in the right panel in Fig.~\ref{fig:1}, 
where the instanton corrections are enough suppressed. 
On the other hand, the trans-Planckian axion decay constant is not achieved in this mirror 
quintic CY. For other CY manifolds with negative small Euler characteristic, 
the large $|K_T|$ will lead to the trans-Planckian axion decay constant around the large volume point 
as discussed in Ref.~\cite{Conlon:2016aea}. 

\subsubsection{Multiple axions}
\label{subsubsec:LCS2}
We next extend the previous analysis to the multi moduli case in a similar step in Sec.~\ref{subsubsec:LCS1}. 
The quantum-corrected K\"ahler potential for multi moduli $T^i$ with $i$ being the number of moduli fields 
is brought into the form~\cite{Hosono:1993qy,Hosono:1994ax},
\begin{align}
K(T,\bar{T}) =-\ln &\biggl[ \frac{1}{6}\kappa_{ijk} (T^i +\bar{T}^i)(T^j +\bar{T}^j)(T^k +\bar{T}^k) 
-\frac{\zeta(3)}{4\pi^3}\chi (\tilde{{\cal M}}) \nonumber\\
&+\frac{2}{(2\pi )^2} \sum_\beta \sum_{n=1}^\infty \frac{d_i n_\beta}{n^2} (T^i+\bar{T}^i) 
e^{-\pi n d_k (T^k+\bar{T}^k)}{\rm cos}\left( -i\pi n d_j (T^j-\bar{T}^j)\right) 
\nonumber\\
&+\frac{4}{(2\pi )^3} \sum_\beta \sum_{n=1}^\infty \frac{n_\beta}{n^3}
e^{-\pi n d_k (T^k+\bar{T}^k)}{\rm cos}\left( -i\pi n d_j (T^j-\bar{T}^j)\right)\biggl],
\label{eq:KmLCS}
\end{align}
around the large volume region, where 
the exponential terms show the instanton corrections labeled by the instanton number $n_\beta$ and 
the degree of internal two-cycles $d_k$. 
From the K\"ahler potential~(\ref{eq:KmLCS}), we find that there are the 
discrete axionic shift symmetries, ${\rm Im}\,T^i\rightarrow {\rm Im}\,T^i +N$ with $N$ being integer. 
Thus, we identify the axions as ${\rm Im}\,T^i$ and the magnitude of these decay 
constant is proportional to $f\propto {\rm det}(K_{T^i\bar{T}^j})$ due to the non-canonical 
kinetic terms of moduli fields. 
The moduli kinetic terms including the instanton correction become 
\begin{align}
K_{T^i\bar{T}^j} =K_{T^i}K_{\bar{T}^j}-e^{K}\biggl[ \kappa_{ijk}(T^k+\bar{T}^k) 
-\sum_\beta \sum_{n=1}^\infty \frac{n_\beta d_id_j}{\pi n} 
e^{-\pi n d_k (T^k+\bar{T}^k)}\biggl],
\end{align}
where
\begin{align}
K_{T^i}=&-e^{K}\Biggl[\frac{\kappa_{ijk}(T^j+\bar{T}^j)(T^k+\bar{T}^k)}{2} 
-\sum_\beta \sum_{n=1}^\infty
\frac{d_id_jn_\beta(T^j+\bar{T}^j)}{2\pi n}e^{-\pi n d_k (T^k+\bar{T}^k)} 
{\rm cos}\left(-i\pi n d_k (T^k-\bar{T}^k)\right)
\nonumber\\
&+i
\sum_\beta \sum_{n=1}^\infty n_\beta \left(
\frac{2d_id_j(T^j+\bar{T}^j)}{2\pi n} +\frac{2d_i}{(2\pi)^2n^2}\right)e^{-\pi n d_k (T^k+\bar{T}^k)} 
{\rm sin}\left(-i\pi n d_k (T^k-\bar{T}^k)\right)\Biggl].
\label{eq:KT2}
\end{align}
When all the moduli fields stay at the points satisfying 
\begin{align}
K_{T^i}=0,
\label{eq:conl1}
\end{align}
or 
\begin{align}
\frac{T^j+\bar{T}^j}{2}K_{T^j} +1&=0,
\nonumber\\
{\rm Im}\,T^i&=0,
\label{eq:conl2}
\end{align}
the moduli K\"ahler metrics satisfies the following equalities,
\begin{align}
\left(T^j+\bar{T}^j\right)K_{T^i\bar{T}^j} =0,
\label{eq:K1}
\end{align}
from which, in both cases, 
the determinant of moduli K\"ahler metric vanishes simultaneously. 
This is because when the inverse matrix of moduli K\"ahler metric exists, 
all the moduli values are equal to zero at the points satisfying Eqs.~(\ref{eq:conl1}) or (\ref{eq:conl2}),
\begin{align}
T^j+\bar{T}^j=0,
\end{align}
which contradict the large volume region of moduli space. 
As a result, at the certain moduli values satisfying Eqs.~(\ref{eq:conl1}) or (\ref{eq:conl2}), 
we obtain the vanishing determinant of moduli K\"ahler metric. 
Although the vanishing inverse K\"ahler metric is unphysical, 
we expect that the tiny determinant of moduli K\"ahler metric, i.e., tiny axion decay constant, 
is realized at the values close to those 
of Eqs.~(\ref{eq:conl1}) or (\ref{eq:conl2}). 


However, when the K\"ahler metric around the points in Eqs.~(\ref{eq:conl1}) or (\ref{eq:conl2}) 
vanishes by the inclusion of  instanton effects, the moduli values are determined by 
the cancellation between the classical and instanton contributions. 
It implies that, in a similar reason discussed in the previous section~\ref{subsubsec:LCS1}, 
these points are not in the large volume region, but in the conifold or 
other regions of special points. 
Thus, for the multiple axion cases, the tiny K\"ahler metric is only achieved around the 
large volume region $T^i\rightarrow \infty$ leading to $f\propto {\rm det}(K_{T^i\bar{T}^j})\rightarrow 0$ in 
which the instanton corrections are enough suppressed. 
On the other hand, the trans-Planckian axion decay 
constant will be realized for certain class of CY manifold with negative small Euler characteristic~\cite{Conlon:2016aea}.

\section{Massless axion and moduli stabilization}
\label{sec:moduli}
As a result of the discussion presented so far, 
we have assumed that the moduli values are treated as parameters to obtain the tiny moduli K\"ahler 
metric, i.e., the tiny axion decay constant. 
In this section, we explore to stabilize the relevant complex structure modulus in type IIB string theory 
with an emphasis on the SCS point as discussed in Sec.~\ref{subsec:scs}. 
To solve the strong CP problem and explain the current dark matter abundance, 
the massless axion with small axion decay constant 
and massive saxion (scalar partner of axion) are required at least at the compactification scale. 
The Planck data also favors the small axion decay constant in order not to 
overproduce the isocurvature perturbation~\cite{Ade:2015lrj}. 
As mentioned before, the couplings between our considered axions and gauge bosons 
are unrevealed for general CY background. Thus, 
in a way similar to the toroidal background~\cite{Lust:2003ky,Blumenhagen:2006ci}, 
we assume that the one-loop gauge kinetic function involves the axion associated 
with complex structure moduli.

To obtain the tiny decay constant of massless axion, 
we consider the moduli stabilization where the saxion $|z|$ is 
stabilized at the minimum satisfying $\partial_z K(|z|)\simeq 0$ 
by introducing the axion-independent uplifting sector 
rather than the superpotential of complex structure modulus $z$. 
In such a case, the axion becomes massless, since the K\"ahler potential is invariant 
under the monodromy transformation around the special point. 
We demonstrate the above scenario at the orbifold point of moduli space 
in Sec.~\ref{subsubsec:orb}, for simplicity. 
For our purpose, we redefine the complex structure modulus 
$(hz)^{-\alpha_1}$ as $\varphi$ and then the K\"ahler potential is simplified as
\begin{align}
K&=-\ln (-i (\tau -\bar{\tau})) +K(T+\bar{T}) +K(|\varphi|),
\nonumber\\
W&=W(\tau, T),
\label{eq:K}
\end{align}
where $\tau$ and $T$ represent the axion-dilaton and K\"ahler moduli 
and $K(|\varphi|)$ corresponds to the K\"ahler potential in Eq.~(\ref{eq:orbK}) 
replaced by $\varphi$,
\begin{align}
K(|\varphi|) =
-\ln
\biggl[ 1 +\frac{C_2}{C_1}|\varphi|^{-2+2\alpha_2/\alpha_1} +{\cal O}(|\varphi|^{-2+2\alpha_3/\alpha_1})
\biggl],
\end{align}
with $C_k=\frac{A_k^2}{4\pi^6} s_k^3c_k\kappa \left( -b s_k^2 +2\left( \frac{2}{3}s_k^2 +c_k^2\right)\right)$.
Now, we perform a gauge transformation of the holomorphic three-form 
to achieve $K(|\varphi|) \rightarrow K(|\varphi|) +\ln |C_1\varphi|^2$ along with Ref.~\cite{Giryavets:2003vd},
\begin{align}
\Omega \rightarrow (C_1\varphi)^{-1}\Omega.
\end{align}
In addition, we assume that the superpotential is non-perturbatively generated by 
the $D$-instanton or gaugino condensation on the hidden D$7$-branes 
wrapping on the internal cycle without depending on the complex structure modulus,
\begin{align}
W&=W(\tau, T),
\label{eq:W}
\end{align}
where we consider the racetrack-type superpotential~\cite{Krasnikov:1987jj,Taylor:1990wr,Casas:1990qi,deCarlos:1992da}. 
When the complex structure modulus appears in the superpotential, the axion 
(phase direction of $\varphi$) generically become massive. 
In this sense, we focus on the $\varphi$-independent superpotential to 
obtain the phenomenologically favorable light axion. 

However, in the current setup, the supersymmetric minimum satisfying 
\begin{align}
D_I W=W_I +K_I W=0,
\label{eq:sta}
\end{align}
where $W_I=\partial W/\partial \Phi^I$ and $K_I =\partial K/\partial \Phi^I$ with $\Phi^I$ 
being the axion-dilaton, K\"ahler and complex structure moduli, is not physical, 
since the K\"ahler metric of complex structure moduli vanishes at this minimum. 
In addition to the K\"ahler and superpotentials in Eqs.~(\ref{eq:K}) and~(\ref{eq:W}), 
we thus introduce the uplifting sector such as anti D$3$-brane~\cite{Kachru:2003aw} 
around the large volume and small complex structure points of CY manifold, 
that is, $T\gg 1$ and $|\varphi|\ll 1$. 
The scalar potential is then characterized by 
the shift symmetric form~\cite{Choi:2004sx,Choi:2005ge},
\begin{align}
V_{\rm up}=e^{2K/3}{\cal P}(T+\bar{T}, |\varphi|)
\end{align}
where ${\cal P}$ is the $T$- and $\varphi$- dependent function. 
The explicit form of ${\cal P}$ depends on the detailed D-brane setup. 

From the scalar potential in four-dimensional ${\cal N}=1$ supergravity,
\begin{align}
V=V_F +V_{\rm up},
\end{align}
with 
\begin{align}
V_F=e^K\left( K^{I\bar{J}}D_IW D_{\bar J} \bar{W} -3|W|^2\right),
\end{align}
we find the local minimum of the complex structure modulus. 
The extremal condition of $\varphi$ is yielded by
\begin{align}
\partial_\varphi V 
\simeq m_{3/2}^2
\biggl[ K_{\bar{\varphi}} \partial_\varphi (K^{\varphi \bar{\varphi}} K_{\varphi})
 +3\partial_\varphi (\ln{\cal P})\biggl]
=0,
\label{eq:Vext}
\end{align}
where $m_{3/2}=e^{\langle K\rangle/2}\langle W\rangle$ is the gravitino mass. 
Now, we employ the condition of almost vanishing cosmological constant $\langle V\rangle \simeq 0$ 
and $\langle V_{\rm up}\rangle \simeq 3m_{3/2}^2$ 
which is valid for the well-known K\"ahler moduli stabilization, e.g., 
Kachru-Kallosh-Linde-Trivedi (KKLT)~\cite{Kachru:2003aw} or racetrack scenario~\cite{Krasnikov:1987jj,Taylor:1990wr,Casas:1990qi,deCarlos:1992da}.

Guided by the K\"ahler potential in Eq.~(\ref{eq:K}), we obtain the following equality,
\begin{align}
 K_{\bar{\varphi}} \partial_\varphi (K^{\varphi \bar{\varphi}} K_{\varphi}) 
 &\simeq -\left(\frac{C_2}{C_1}\right)^2 \left(\frac{\alpha_2}{\alpha_1}-1\right) 
 |\varphi|^{4\left(\frac{\alpha_2}{\alpha_1}-1\right) }\varphi^{-1},
\end{align}
and consequently the vacuum expectation value of complex structure modulus 
is determined to satisfy the extremal condition~(\ref{eq:Vext}),
\begin{align}
|\varphi|^{4\frac{\alpha_2}{\alpha_1}-5}&\simeq 
3\left(\frac{C_1}{C_2}\right)^2\left(\frac{\alpha_2}{\alpha_1}-1\right) 
|\partial_\varphi (\ln{\cal P})|,
\label{eq:VEV}
\end{align}
where $\alpha_2/\alpha_1$ is grater than unity in our examples in 
Tab.~\ref{tab1} with distinct $\alpha_i$, and the saxion $|\varphi|$ is only stabilized at the 
above minimum by the axion-independent 
scalar potential. 
Now, we assume that the mass squared of saxion is positive, 
since they depend on the second derivative of $\partial_\varphi\partial_{\bar \varphi}(\ln {\cal P})$. 
The other closed string moduli can be stabilized at the vacuum by the 
superpotential. 


In this way, it is possible to obtain the massless axion along the above moduli stabilization procedure. 
From the formula of axion K\"ahler metric,
\begin{align}
K_{\theta\theta}=K_{\varphi{\bar \varphi}}|\varphi|^2
\simeq -\left( \frac{\alpha_2}{\alpha_1}-1\right)^2\left(\frac{C_2}{C_1}\right)
|\varphi|^{2\left (\frac{\alpha_2}{\alpha_1}-1\right)},
\end{align}
and the vacuum expectation value of $|\varphi|$ in Eq.~(\ref{eq:VEV}), 
the tiny K\"ahler metric of axion $\theta={\rm arg}(\varphi)$, i.e., the tiny axion decay constant, 
is achieved only when $\partial_\varphi (\ln{\cal P})$ is much smaller than unity. 
For example, the cosmologically favorable axion K\"ahler metric $\sqrt{2K_{\theta\theta}}\simeq 10^{12}\,{\rm GeV}$ 
is achieved under $\partial_\varphi (\ln{\cal P}) \simeq 3\times 10^{-7}$ for the CY manifold 
defined in $\mathbb{P}^5_{3,2,2,1,1,1}[4,6]$. 
In a similar fashion, we can realize the suppressed axion decay constant at 
the other SCS points and conifold point. 
Note that the K\"ahler metric is governed by the vacuum expectation value of $\partial_\varphi (\ln{\cal P})=\partial_{\varphi} {\cal P}/{\cal P}$, 
whereas the cosmological constant is dominated by that of ${\cal P}$. 
Therefore, the obtained small axion K\"ahler metric is compatible with the tiny cosmological constant under 
$\partial_{\varphi} {\cal P} \ll {\cal P}$. 
However, it is interesting and challenging issue to obtain such a small value of $\partial_\varphi (\ln{\cal P})$, 
since its specific form depends on the concrete D-brane setup taking into account the 
Ramond-Ramond tadpole condition. 
We study the detailed construction of tiny axion decay constant elsewhere.


\section{Conclusion}
\label{sec:con}

In this paper, we have studied the detail of the quantum and 
geometrical corrections for a decay constant of closed string axion with an 
emphasis on the special points of Picard-Fuchs equation, 
in particular the small complex structure 
point in Sec.~\ref{subsec:scs}, conifold point in Sec.~\ref{subsec:coni}, 
and large complex structure (large volume) point in Sec.~\ref{subsec:LCS}, 
around which the axion particles naturally appear in the 
low-energy effective theory. 
The axionic shift symmetries of the K\"ahler potential are originating from 
the geometrical symmetries around the special points. 
Furthermore, in the type IIB string theory, the decay constant of axion associated with the complex structure 
modulus is irrelevant with the string scale in comparison with that of K\"ahler 
modulus. 
It is thus interesting to discuss the axion associated with complex structure modulus. 

On the basis of topological string theory, we find the general expression of 
period vector and its monodromy transformation for typical one-parameter CY manifolds 
in Tab.~\ref{tab1} by solving the corresponding Picard-Fuchs differential equations. 
It then turns out that only the tiny axion decay constant can be 
realized around the special point, since the axion K\"ahler metric exactly vanishes at 
all the special points. 
We also demonstrate the moduli stabilization to realize such 
a situation in Sec.~\ref{sec:moduli}. It is interesting to embed our scenario in 
the detailed D-brane setup and reveal the couplings between the axion and 
matter fields in the visible sector, which will be studied elsewhere.

\section*{Acknowledgments}
The authors thank H.~Abe for valuable comments. 
H.~O. would like to thank K.~Choi, R.~Gopakumar, T.~Higaki and G.~Shiu for useful discussions. 
A.~O. and H.~O. are grateful to Y.~Honma for fruitful discussions. 
A.~O. and H.~O. also thank the Yukawa Institute for Theoretical Physics at Kyoto University. 
Discussions during the YITP workshop YITP-W-16-05 "Strings and Fields 2016" 
were useful to complete this work.

\bibliography{ref}



\end{document}